\documentclass[preprint,12pt,authoryear]{elsarticle}



\usepackage{natbib}
\usepackage{amsmath}
\usepackage{txfonts}
\usepackage{amssymb}

\usepackage{lipsum} 
\usepackage[T1]{fontenc}
\usepackage{stfloats}
\usepackage{graphicx}
\usepackage[normalem]{ulem}

\usepackage[table,xcdraw]{xcolor}
\usepackage{latexsym,mathrsfs}
\usepackage[hidelinks]{hyperref}
\usepackage{float}
\usepackage{epsfig}
\usepackage{soul}
\usepackage{makecell}
\usepackage{colortbl} 
\usepackage{longtable}
\usepackage{adjustbox}
\usepackage{multirow}
\usepackage{booktabs}
\usepackage{bookmark}
\usepackage{siunitx}
\usepackage{subcaption}

\journal{High Energy Astrophysics}

\begin{document}

\begin{frontmatter}

\title{On the Metric \texorpdfstring{$f(R)$}{f(R)} gravity Viability in Accounting for the
Binned Supernovae Data}

\author[1,2]{A. Valletta}{}
\cortext[cor1]{Corresponding author}
\ead{andrea.valletta@uniroma1.it}

\affiliation[1]{organization={Physics Department, Sapienza University of Rome},
            addressline={P.le A. Moro 5}, 
            city={Rome},
            postcode={00185}, 
            country={Italy}}
            
\affiliation[2]{organization={Istituto Nazionale di Fisica Nucleare (INFN), Sezione di Roma},
            addressline={P.le A. Moro 5}, 
            city={Rome},
            postcode={00185}, 
            country={Italy}}
            
\author[3,1]{G. Montani}{}
\ead{giovanni.montani@enea.it}
\affiliation[3]{organization={ENEA, Nuclear Department, C.R. Frascati},
            addressline={Via E. Fermi 45}, 
            city={Frascati},
            postcode={00044}, 
            country={Italy}}

\author[4,5,6,7]{M. G. Dainotti}{}
\ead{maria.dainotti@nao.ac.jp}
\affiliation[4]{organization={Division of Science, National Astronomical Observatory of Japan},
            addressline={2 Chome-21-1 Osawa, Mitaka}, 
            city={Tokyo},
            postcode={181-8588}, 
            country={Japan}}
            
\affiliation[5]{organization={Astronomy Department, The Graduate University for Advanced Studies, SOKENDAI},
            addressline={Shonankokusaimura, Hayama, Miura District}, 
            city={Kanagawa},
            postcode={240-0115}, 
            country={Japan}}
\affiliation[6]{organization={Space Science Institutes },
            addressline={4765 Walnut St Ste B}, 
            city={Boulder},
            postcode={ 80301}, 
            state={CO},
            country={USA}}
            
\affiliation[7]{organization={Nevada Center for Astrophysics, University of Nevada,89154},
            addressline={4505 Maryland Parkway}, 
            city={Las Vegas},
            postcode={ 80301}, 
            state={NV},
            country={USA}}

\author[1,2,8]{E. Fazzari}{}
\ead{elisa.fazzari@uniroma1.it}

\affiliation[8]{organization={Physics Department, Tor Vergata University of Rome},
            addressline={Via della Ricerca Scientifica 1},
            postcode={00133}, 
            country={Italy}}

\begin{abstract}
In this work, two models of metric $f(R)$ gravity in the Jordan frame are investigated as a dynamical description of the late-time cosmic expansion using binned Type~Ia Supernovae data.
The aim is to provide an explanation for the effective running of the Hubble constant observed in both the binned Pantheon Sample and the Master Sample. To this end, the effective running Hubble constant $\mathcal{H}(z)$ is defined as the ratio between the modified Hubble parameter and that of the standard cosmological model ($\Lambda$CDM), multiplied by $H_0$. $\mathcal{H}(z)$ serves as a diagnostic tool to capture deviations from the $\Lambda$CDM model.

The first model used is a general representation of metric $f(R)$ gravity in which the gravitational Lagrangian is encoded in an effective redshift-dependent function that mimics the evolution of the Hubble parameter. This function can be reliably approximated by a second-order Taylor expansion at low redshift.

While this more general formulation yields a phenomenological fit that is better than the $\Lambda$CDM for the binned Pantheon Sample, the model generically leads to the emergence of an unphysical mass of the scalar field. This issue originates from an implicit restriction imposed on the Cauchy problem for the non-minimally coupled scalar field. To address this limitation, following previous studies, an additional condition on the modified Friedmann equation is introduced, enabling a fully consistent reformulation of the dynamics. 
It is clarified that the additional condition has a precise dynamical origin, being necessary to restore a consistent Cauchy problem and to ensure a finite, positive scalar field mass.
The resulting framework not only preserves the agreement with binned Supernova~Ia data, but also provides a physical justification for the additional condition adopted in earlier analyses of late-time cosmological dynamics.
\end{abstract}

\end{frontmatter}
\section{Introduction}\label{Intro}
Recent analyses of binned Type Ia Supernovae (SNe~Ia) data have revealed that the
inferred value of the Hubble constant exhibits a decreasing trend as the
redshift increases, a behaviour consistently reported in several works
\citep{dainotti2021hubble, kazantzidis2020, cinesi_H0z,  dainotti2022evolution, dainotti2025new}.  
This trend has been clearly identified both in the binned Pantheon sample
\citep{dainotti2021hubble, Scolnic2018} and in the so-called ``Master sample'', a
homogeneous dataset constructed by combining multiple SNe~Ia compilations
\citep{dainotti2025new}.
In this context, the binning approach adopted in the analysis of the Master Sample plays a crucial role from a statistical point of view.
Rather than assuming \emph{a priori} a Gaussian likelihood for the residuals between observed and theoretical distance moduli, the likelihood function is explicitly tested in each redshift bin.
As shown in \citet{dainotti2024statistical}, the distribution of the residuals normalised by the full covariance matrix departs significantly from a Gaussian form in several SNe~Ia samples.
In particular, for Pantheon+ and Pantheon data the best-fit likelihoods are found to be better described by heavy-tailed distributions, such as a Student-$t$ or a logistic distribution, respectively.

This procedure makes the underlying statistical assumptions explicit and avoids imposing an incorrect Gaussian likelihood where it is not supported by the data.
As a consequence, the inferred values of $H_0$ and their uncertainties in each redshift bin are obtained using the statistically preferred likelihood, leading to a more robust reconstruction of the redshift dependence of the effective Hubble constant.

For similar analyses applied to Gamma-Ray Bursts, see
\citet{dainotti2022gamma,dainotti2013determination,dainotti2022optical,dainotti2013slope,dainotti2015luminosity,dainotti2017study,dainotti2023gamma};
for quasars, see
\citet{dainotti2024new,dainotti2023quasars,dainotti2022quasar,lenart2023bias};
for gravitational waves, see
\citet{su2025exploring,pan2025determining,zhan2025precise};
and for other cosmological probes, see
\citet{postnikov2014nonparametric,adil2024revisiting,mukherjee2025constraining,chen2025measuring}. 
The most reliable analytical representation of $H_0$ follows a power law behaviour, which has been interpreted either as a possible astrophysical evolution in
SNe~Ia standardisation or as an indirect indication of departures from the
standard $\Lambda$CDM framework. 
The decreasing behaviour has also been confirmed by DESI through a BAO analysis in \citet{jia2025uncorrelated, Jia2025} (see also \citet{Xu:2026gdq}, \citet{Wu:2026svr} and \citet{Chaudhary:2025vzy}). 

The possibility that this behaviour originates from a modified expansion
history has been investigated through the diagnostic tool known as the
effective running Hubble constant, defined as the ratio between the
Hubble parameter of a given cosmological model and that predicted by
$\Lambda$CDM, all multiplied by $H_0$ 
\citep{dainotti2021hubble,schiavone2023f,montani2025running} (for a similar investigation concerning the effective matter critical parameter, see \citet{dainotti2024new}).  
Previous studies \citep{schiavone2023f,montani2025two} have shown that a
power law decay of this diagnostic tool can be explained within metric
$f(R)$ gravity or interacting dark-energy-dark-matter models.  
Other physically motivated formulations have been
proposed in recent literature
\citep{montani2024slow,montani2025running,montani2025decay,fazzari2025effective, cianfrani2014canonical, giare2025overview, giare2025dynamical,wang2025can,wang2025evidence, giare2024interacting,silva2025new,zhai2025low,kessler2025one,manoharan2025solves,dixit2025beyond, giare2024robust,fazzari2025cosmographic}, see also \citet{kalita2025revealing}.
For multi-messenger tests of modified gravity see \citet{zhang2021distinguish,su2025einstein,zhang2026standard, Gadbail:2026hqo}.

The power law decrease of the effective running Hubble constant has also
been linked to the so-called ``Hubble tension''
\citep{di2021realm,di2025cosmoverse, dainotti2024reduceduncertainties43hubble}, namely the $\sim4\sigma$ discrepancy
between the value of $H_0$ inferred from cosmic microwave background
measurements (e.g.\ by \emph{Planck} \citep{Planck2018}) and that obtained
from the Cepheid-calibrated SNe~Ia distance ladder (e.g.\ by the SH0ES
Collaboration \citep{Riess2022}). For other measurenments of the Hubble constant, see also \citet{gadbail2026model,verma2026stellar}.
This connection has been explored from both observational
\citep{dainotti2021hubble,dainotti2025new,bochner2026exploring} and theoretical perspectives
\citep{montani2024metric,courbin2026hubble}, including within modified gravity frameworks \citep{Ladeira:2026jne, Oliveira:2025uye}. Other constraints on the cosmological parameters can be found in \citet{Pardo:2026apx} and \citet{Zhou:2025dxo}.
Among these, metric $f(R)$ gravity in the Jordan frame has received
particular attention, especially in formulations where an additional
condition is imposed on the Friedmann equation to treat the scalar field
potential as a genuine dynamical variable \citep{SotiriouFaraoni2010}.  
Such a prescription yields a consistent dynamical scheme capable of
alleviating the Hubble tension, although its physical origin has remained
partly ambiguous. Other studies regarding the Hubble tension can be found in \citet{lohakare2026hubble,Yarahmadi2026,Yashiki2025,goswami2025constraining,carloni2025addressing,lee2025alleviating,gurzadyan2025cosmic,lee2025geometric,lenart2023bias,de2025doublet,ling2025model,desmond2025subtle,leclair2025quantum, Montani:2026owo}.
Effects of dark matter and bulk viscosity effect can be found in \citet{brevik2017viscous, brevik2011viscous, belinskii1975influence, capozziello2006observational, nojiri2005inhomogeneous, belinskii1977viscosity, belinskii1979investigation, montani2017bianchi, carlevaro2008gravitational, disconzi2015new, Oikonomou:2025qub}.

The present work aims to provide a physical and dynamical explanation of the additional condition imposed on the Friedmann equations, clarifying its role in restoring a well-posed Cauchy problem and ensuring the viability of metric $f(R)$ gravity at late times.

The analysis is carried out by adopting a general reconstruction scheme for
metric $f(R)$ gravity in the Jordan frame, in which the gravitational
Lagrangian is parametrized through an effective redshift dependent function
that governs the evolution of the Hubble parameter.
Because the binned SNe~Ia data probe only the late-time, low redshift Universe,
this function can be expanded around the present epoch with
a second-order Taylor series. This provides a
parametrisation that is sufficiently flexible to capture the observed
dynamics while remaining statistically stable and well controlled.

Despite obtaining a good phenomenological agreement with SNe~Ia
observations, this reconstruction is found to be physically inconsistent.
In particular, the resulting theory generically develops a tachyonic
instability in the scalar degree of freedom.
This behaviour can be traced back to the reconstruction procedure itself,
which implicitly over-constrains the Cauchy problem for the non-minimally
coupled scalar field.
Indeed, both the present-day value of the scalar field and its time
derivative are fixed simultaneously at $z=0$, despite the fact that the scalar
equation of motion is second-order.
Such an over-constraint is incompatible with a consistent dynamical evolution
and inevitably leads to pathological behaviour in the scalar sector.

To address this issue, the analysis is reformulated within the dynamical
framework introduced in \citet{montani2024metric}, where an additional condition
is imposed on the modified Friedmann equation.
This prescription removes the spurious restriction on the initial data,
allowing the present-day time derivative of the scalar field to be non zero.
As a consequence, the mass of the scalar field remains finite and positive, and the
resulting theory becomes physically viable.

Within this revised framework, metric $f(R)$ gravity in the Jordan frame
\citep{capozziello2011extended} is studied by encoding the form of the
gravitational Lagrangian in a redshift dependent function that effectively
rescales the model Hubble parameter.
The Chevallier-Polarski-Linder (CPL) parametrisation \citep{Chevallier2001, Linder2003} is
adopted to describe the evolution of the dark energy component.
This choice is motivated by recent results from the DESI collaboration \citep{adame2025desi,karim2025desi, lodha2025desi}, which,
through analyses of baryon acoustic oscillations, indicate a preference for
an evolving dark energy scenario over the standard $\Lambda$CDM model (for other probes of a departure from the standard cosmological model, see \citet{Scherer:2025esj, Das:2023hbw, Das:2013sca}).

Within the DESI's CPL framework, dark energy undergoes a transition from a
quintessence-like regime to a phantom-like behaviour.
This feature is consistent with results obtained from non-parametric
reconstructions of the dark energy equation of state
\citep{Berti2025,Nesseris2012,gonzalez2025reconstruction,Lodha2025,efstratiou2025addressing,
linder2024interpreting,Dymnikova1998,Dymnikova2000,Dymnikova2001,
Doroshkevich1984,Doroshkevich1985,Doroshkevich1988}, as well as from
model independent approaches \citep{fazzari2025cosmographic}.

In this consistent dynamical setting, the model continues to be statistically in agreement with both the binned Pantheon and Master SNe~Ia samples, while avoiding
tachyonic instabilities and divergences of the scalar field mass.
These results therefore supply a clear physical justification for the
additional condition introduced in previous studies of late-time
cosmological dynamics.

\medskip

The paper is organised as follows.  
In Sec.~\ref{Data} the two binned SNe~Ia datasets used in this work, the
Pantheon sample and the Master sample, are introduced.  
In Sec.~\ref{gravity} metric $f(R)$ gravity in the Jordan frame is
presented, including the action, the field equations, the equivalent
scalar-tensor formulation, and the definition and physical role of the
scalar field mass.  
The resulting modifications to the Friedmann equation are also discussed.  
The next two sections contain parallel analyses.  
Sec.~\ref{phenomenology} examines the first model, based on a phenomenological
function $f(z)$: the model is introduced, tested against the binned Pantheon
sample, then against the binned Master sample, and the main results are
summarised.  
Sec.~\ref{sec:newmodel} introduces the second $f(R)$ model, featuring the
additional condition on the potential; it is tested on both datasets, and
general conclusions are extracted.  
Finally, Sec.~\ref{conclusions} summarises the main results and outlines
possible future developments.

\section{Binned Supernovae Data}
\label{Data}

In this work, two modified gravity models of the metric $f(R)$ type are tested against binned SNe~Ia datasets. Specifically, the analysis employs the binned Pantheon Sample and the Master Sample.

It is worth stressing that the Type~Ia supernova samples considered in this work are not uniformly calibrated through a Cepheid-based distance ladder.
With the exception of specific low-redshift subsets of Pantheon+, the Pantheon, JLA, and DES compilations are constructed as relative distance indicators.
In these datasets, the absolute magnitude of SNe~Ia is treated as a nuisance parameter and marginalised over in cosmological analyses, so that no absolute determination of $H_{0}$ is obtained without the inclusion of external information or priors (e.g. from BAO or CMB data).
Only dedicated local analyses of nearby Pantheon+ subsamples explicitly adopt a Cepheid calibration to infer a local value of the Hubble constant.
As a consequence, the redshift dependence of the inferred $H_{0}(z)$ discussed in this work cannot be trivially attributed to a specific local calibration method, such as Cepheids, but instead emerges consistently across independently constructed SN~Ia samples with different calibration strategies.

\subsection{The Pantheon Sample}
\label{Pantheon}

The Pantheon Sample is one of the largest and most homogeneous compilations of SNe~Ia to date, consisting of 1048 spectroscopically confirmed events collected from a variety of surveys. When calibrated through the distance-ladder method, SNe~Ia act as standardisable candles: by combining measurements of their luminosity distance with their redshift, it becomes possible to infer an individual estimate of the Hubble constant $H_{0}$ for each event.

The binned Pantheon sample is considered here instead of Pantheon+ (see \citet{chandak2026model}), since the latter is already included in the Master Sample (with a contribution significantly larger than the former). Studying Pantheon separately allows a direct comparison with previous analyses testing $f(R)$ gravity models \citep{dainotti2021hubble, dainotti2022evolution, montani2025two}.

For a complementary analysis of the same Pantheon sample binned through an MCMC procedure with free $H_{0}$ and a Gaussian prior on $\Omega_{m}^{0}$, leading to statistical properties fully consistent with those of the present binned configuration, see \citet{dainotti2022evolution}.

To ensure internal consistency, the Pantheon dataset includes several corrections, such as Milky Way extinction, light-curve colour and stretch, host galaxy mass, selection effects, and both statistical and systematic uncertainties, as detailed in \citet{Scolnic2018} (see also \citet{dainotti2023reducing,dainotti2024statistical}).

To explore potential redshift evolution in the Hubble parameter, the full dataset is subdivided into redshift bins. For each equipopulated bin, an independent estimate of $H_{0}$ is obtained, and the full profile $H_{0}(z)$ is reconstructed by combining the values across all bins. 

In this analysis, we adopt the configuration of 40 bins previously examined in \citet{dainotti2021hubble,dainotti2022evolution}. Each bin contains approximately 26 SNe~Ia, except for the final bin, which includes 34 events. For comparison, the classical study of \citet{Perlmutter1999} employed only 42 SNe~Ia in total. Despite the modest number of objects per bin, the 40-bin configuration yields stable and consistent results: both the inferred value of the Hubble constant and the absolute magnitude remain in agreement, within uncertainties, with those obtained from alternative binning approaches.

The absolute luminosity calibration is fixed by anchoring the first redshift bin to $H_{0}=73.5\,\mathrm{km\,s^{-1}\,Mpc^{-1}}$. This yields an absolute magnitude $M=-19.250 \pm 0.021$, which is subsequently held fixed in all remaining bins. The matter density parameter is also fixed to $\Omega_{m}=0.298$, consistent with previous analyses.

This calibration strategy reflects the intrinsic degeneracy between the absolute magnitude $M$ and the Hubble constant $H_{0}$ in Type~Ia supernova analyses. 
Fixing $H_{0}$ in the first redshift bin allows a reference value of $M$ to be determined, which is then consistently propagated to all remaining bins, enabling the reconstruction of relative variations of $H_{0}(z)$ without introducing additional degeneracies.

Earlier studies \citep{dainotti2021hubble,dainotti2022evolution} report a mild yet persistent decrease of $H_{0}$ with increasing redshift. The present analysis confirms this behaviour within the 40-bin configuration , consistent with analogous results obtained using 3, 4, or 20 bins. Moreover, the estimates of the absolute magnitude $M$ across all binning schemes remain compatible at the $1\sigma$ level.

\subsection{The Master Sample}
\label{sec:master_sample}

A key component of the present investigation of the effective running Hubble constant is the Master Supernovae Ia (SNe~Ia) Sample introduced in \citet{dainotti2025new}. This dataset is constructed by merging the major existing SNe~Ia catalogs into a single homogeneous compilation, free from duplicate entries. It combines four major surveys: Pantheon \citep{Scolnic2018}, Pantheon+ (P+) \citep{Scolnic_2022, Brout_2022}, the Joint Lightcurve Analysis (JLA) \citep{Betoule2014} and the Dark Energy Survey (DES) \citep{DES2019}. After a rigorous cross-matching procedure and the removal of repeated events, the final sample contains 3714 unique SNe~Ia, substantially larger than any individual catalog.

The construction of the Master Sample follows a strict protocol. Catalogs are prioritised in reverse chronological order: DES, P+, Pantheon, and finally JLA, so that the most recent observations take precedence when duplicates occur. DES, with 1829 spectroscopically confirmed SNe~Ia, contributes the largest subset. Pantheon+ initially contained 1701 events, but 158 internal duplicates and 335 overlaps with DES reduce its net contribution to 1208 SNe~Ia. Pantheon, originally comprising 1048 SNe~Ia, contributes 181 unique objects, while JLA adds 496 events after accounting for overlaps.

A distinctive feature of the Master Sample is its calibration strategy. Initially, the absolute magnitudes and $H_{0}$ calibrations of each subsample are preserved to avoid introducing artificial scatter. Subsequently, a uniform recalibration of the absolute magnitude parameter $M$ is performed across the entire dataset, ensuring consistent normalisation while preserving any intrinsic redshift evolution in $H_{0}(z)$. In addition, the covariance matrices from each catalog, including both statistical and systematic uncertainties, are merged to ensure a consistent propagation of errors.

In practice, this uniform recalibration is implemented by applying a constant offset to the distance moduli, corresponding to a shift in the absolute magnitude parameter $M_B$ associated with the adoption of the fiducial value $H_0 = 70\,\mathrm{km\,s^{-1}\,Mpc^{-1}}$.
This procedure amounts to subtracting a constant magnitude offset from the observed distance moduli, while preserving the original light-curve fit parameters and covariance matrices of each subsample.

It is important to clarify that the construction of the Master Sample does not imply the adoption of a SH0ES or Cepheid-based calibration across all subsamples.
All datasets entering the Master compilation (Pantheon, Pantheon+, JLA, and DES) are treated as relative distance indicators, and their original light-curve fits and internal calibration procedures are preserved.
No new joint fitting of the supernova light curves using a common pipeline is performed.
The value of the Hubble constant adopted in the individual catalogs (typically $H_0 = 70\,\mathrm{km\,s^{-1}\,Mpc^{-1}}$ for JLA, DES, and Pantheon) is employed purely as a fiducial reference normalization and does not correspond to imposing an external distance-ladder calibration.
To combine the different datasets consistently and avoid artificial offsets, a uniform recalibration of the absolute magnitude parameter $M$ is applied, bringing the Pantheon+ subsample to the same fiducial reference value.
This procedure does not affect relative distance measurements nor the reconstruction of redshift-dependent trends.
In particular, the Master Sample could in principle be constructed even without specifying a value of $H_0$, since the redshift evolution discussed in this work relies on relative variations rather than on an absolute calibration.
The fiducial choice of $H_0$ therefore serves exclusively comparison and consistency purposes and does not bias the results toward a specific local determination such as SH0ES.
A full reanalysis of the light curves with a unified pipeline is beyond the scope of the present work and will be addressed in a forthcoming study.

This comprehensive and statistically powerful dataset enables multiple binning strategies. The case of equi-populated binning is explored in \citet{dainotti2025new}, as well as the case of equally spaced bins in $\log z$ and that of a moving window procedure. These techniques allow us the reconstruction of $H_{0}(z)$ across the full redshift range of the Master Sample, from $z\simeq 0.001$ up to $z\simeq 2.9$. Any observed redshift evolution may indicate astrophysical effects, including SNe~Ia evolution or selection biases, or genuine deviations from standard cosmology.

For part of the present study, the Master Sample is used as the reference catalog to test the theoretical model through a Markov Chain Monte Carlo (MCMC) inference method implemented with tools such as \textsc{Cobaya}. Unlike in the Pantheon analysis, the Master Sample allows for simultaneous constraints on both $\Omega_{m0}$ and $H_{0}$. Moreover, the Master Sample analysis employs the best-fit statistical procedure, following \citet{dainotti2024statistical}, within each redshift bin, ensuring that the assumption of a Gaussian likelihood does not introduce biases in the results.

The Master data set is divided into 20 equally populated redshift bins spanning $0.0012 \leq z \leq 2.3$. The mean redshift of the first bin is $\langle z \rangle \simeq 0.0091$, while the last bin reaches $\langle z \rangle \simeq 1.54$. Given this dataset, each redshift bin is fitted assuming a $\Lambda$CDM model, with $H_0$ and $\Omega_{m0}$ treated as free parameters. The Hubble constant is sampled using a flat prior, while the matter density parameter is sampled using a Gaussian prior, reflecting external constraints from independent cosmological probes. This choice accounts for the limited constraining power of Type Ia supernovae on $\Omega_{m0}$, as discussed in \citet{dainotti2025new}.

\section{\texorpdfstring{Metric $f(R)$ gravity}{Metric f(R) gravity}}
\label{gravity}

Metric $f(R)$ gravity is a generalisation of the Hilbert-Einstein theory in which the Ricci scalar is replaced by a general function of $R$. In the Jordan frame, the action takes the form
\begin{equation}
S = \frac{1}{2\kappa} \int d^4x \, \sqrt{-g} \, f(R) + S_m[g_{\mu\nu}, \psi],
\label{f(R)action}
\end{equation}
where $\kappa = 8\pi G$ (with units $c=1$), $G$ is Newton’s constant, $g$ is the determinant of the metric $g_{\mu\nu}$, and $S_m$ is the matter action depending on the metric and the matter fields $\psi$.  

Varying Eq.~\eqref{f(R)action} with respect to the metric yields the field equations
\begin{equation}
f_R(R) R_{\mu\nu} - \frac{1}{2} f(R) g_{\mu\nu} + 
\left( g_{\mu\nu} \Box - \nabla_\mu \nabla_\nu \right) f_R(R)
= \kappa T_{\mu\nu},
\label{eomf(R)}
\end{equation}
where $f_R(R) \equiv df/dR$, $\Box \equiv \nabla^\alpha \nabla_\alpha$ is the d'Alembertian operator, $R_{\mu\nu}$ is the Ricci tensor, and $T_{\mu\nu}$ is the matter energy-momentum tensor.

\medskip

Metric $f(R)$ gravity is dynamically equivalent to a scalar-tensor theory with a scalar potential \citep{SotiriouFaraoni2010}.  
Introducing the scalar field
\begin{equation}
\phi \equiv f_R ,
\label{phi}
\end{equation}
the action can be rewritten as
\begin{equation}
S_{\text{BD}} = \frac{1}{2\kappa} \int d^4x \, \sqrt{-g}
\left[ \phi R - U(\phi) \right] + S_m[g_{\mu\nu},\psi],
\label{phiaction}
\end{equation}
where the potential satisfies
\begin{equation}
\frac{dU}{d\phi} = R.
\label{R}
\end{equation}

\noindent Varying Eq.~\eqref{phiaction} with respect to the metric and the scalar field gives the field equations
\begin{align}
\phi G_{\mu\nu} &= \kappa T_{\mu\nu}
 + \nabla_\mu \nabla_\nu \phi 
 - g_{\mu\nu} \Box \phi
 - \frac{1}{2} g_{\mu\nu} U(\phi),
 \label{eomphi1}
 \end{align}
\begin{align}
\Box \phi &= \frac{1}{3} \left( 
\kappa T + \phi \frac{dU}{d\phi} - 2 U(\phi)
\right),
\label{eomphi2}
\end{align}
where $T = g^{\mu\nu}T_{\mu\nu}$ is the trace of the matter energy-momentum tensor and $G_{\mu \nu}\equiv R_{\mu \nu}-\frac{1}{2}g_{\mu \nu}R$ is the Einstein tensor.

\subsection{Mass of the scalar field}
\label{Mass}

Metric $f(R)$ gravity introduces an additional scalar degree of freedom, which significantly modifies the cosmological dynamics. Linearizing around a background solution allows one to define an effective mass for this scalar mode (see \citet{SotiriouFaraoni2010}):
\begin{equation}
u^2
= \frac{1}{3} \left( \frac{f_R(R)}{f_{RR}(R)} - R \right),
\label{mass}
\end{equation}
where $f_{RR}(R) = d^2 f/dR^2$.

An effective Newton constant can be obtained from Eq.~\eqref{eomphi1} dividing by $\phi$, so one obtains:
\begin{equation}
    G_{\mu \nu}=8\pi \frac{G}{\phi}T_{\mu \nu}+...=8\pi G_{eff}\cdot T_{\mu \nu}+...
\end{equation}
To ensure a positive effective Newton constant $G_{eff}\equiv\frac{G}{\phi}$, one must impose $\phi>0$, i.e. $f_R > 0$. 
As a consequence, if $f_{RR} < 0$ the effective mass squared becomes negative and the scalaron experiences tachyonic growth, making the theory unstable in that regime.
In this sense, a necessary condition for avoiding tachyonic instabilities is therefore $f_{RR}>0$.

\subsection{\texorpdfstring{Friedmann equations in metric $f(R)$ gravity}{Friedmann equations in metric f(R) gravity}}

For a spatially flat FLRW metric \citep{friedman1922krummung, lemaitre1927univers, robertson1935kinematics, walker1937milne},
\begin{equation}
ds^2 = -dt^2 + a^2(t)(dx^2 + dy^2 + dz^2),
\label{metric element}
\end{equation}
the Ricci scalar is given by 
\begin{equation}
R = 6\left( 2H^2 + \dot{H} \right),
\label{ricci}
\end{equation}
as in \citet{schiavone2023f}.
In Eq.~\eqref{ricci}, $H = \dot{a}/a$ is the Hubble parameter and the dot refers to a differentiation with respect to synchronous time.  

In this background, the modified Friedmann equation reads
\begin{equation}
H^2 = \frac{1}{3\phi}
\left[
\kappa \rho_m+\kappa\rho_{DE}
+ \frac{1}{2} U(\phi)
- 3H \dot{\phi}
\right].
\label{friedmann}
\end{equation}
where $\rho_m$ is the matter density and $\rho_{DE}$ describes the density of dark energy in the universe.

\section{\texorpdfstring{$f(R)$ Gravity from a Phenomenological $f(z)$}{f(R) Gravity from a Phenomenological f(z)}}
\label{phenomenology}

In this section, a phenomenological $f(R)$ gravity model is introduced. It is described starting from a function of the redshift $f(z)$, and it is tested against the binned Pantheon Sample. Finally, the same study is extended to the Master Sample within a Bayesian framework, and some remarks on this first $f(R)$ model are drawn.

\subsection{Model Description}
\label{model descr1}

Using Eqs.~\eqref{ricci},~\eqref{friedmann}, one obtains the following set of equations in terms of the redshift:

\begin{equation}
\begin{split}
&E^2(z) \equiv  \frac{H^2(z)}{H_0^2}\\
&= \frac{1}{f(z)} \Big[
      \Omega_{m0}(1+z)^3 
      + \Omega_{DE}(z)
      + \bar{U}(z)
\Big],
\label{eq finale}
\end{split}
\end{equation}

\begin{equation}
\bar{U}' = \phi'
\left( 
2 E^2 - \frac{1}{2} (1+z) (E^2)'
\right),
\label{evoluzione Ubar}
\end{equation}
where
\begin{equation}
\phi - (1+z)\phi' \equiv f(z),
\label{problemacauchy}
\end{equation}
and the prime $'$ denotes differentiation with respect to $z$, $\bar{U} \equiv U/(6H_0^2)$, and $\Omega_{DE}\equiv \kappa \rho_{DE}/(3H_0^2)$.  
Normalization is fixed by imposing $E(0)=1$ by definition, which requires
\begin{equation}
f(0) = 1 .
\end{equation}

As the dark energy component, a dynamical dark energy parametrization is adopted, specifically the CPL form \citep{Chevallier2001,Linder2003}. In this framework, the dark energy equation of state evolves with redshift as 
\begin{equation}
    \omega(z) = \omega_0 + \omega_a \frac{z}{1+z},
    \label{EOS}
\end{equation}
where $\omega_0$ is the present-day value of the dark energy and $\omega_a$ quantifies its redshift evolution. 

The corresponding dark energy density evolves according to
\begin{equation}
\Omega_{DE}(z)
= (1 - \Omega_{m}^0-\bar{U}_0)\,(1+z)^{3(1+\omega_0+\omega_a)}
  \exp\!\left[ -3\omega_a\frac{z}{1+z} \right],
\label{OmegaDE}
\end{equation}
where $\Omega_{m}^0 \equiv \kappa \rho_{m0} / (3 H_0^2)$ is the present-day matter density parameter and $\bar{U}_0=\bar{U}(z=0)$. 

Eq.~\eqref{problemacauchy} admits the general solution
\begin{equation}
\phi(z) = (1+z)
\left[
A - \int \! dz \, \frac{f(z)}{1+z}
\right],
\label{scalarfield}
\end{equation}
where $A$ is an integration constant. Consistency with General Relativity today requires
\begin{equation}
\phi(z=0) = 1 .
\end{equation}

Substituting Eq.~\eqref{eq finale} into Eq.~\eqref{evoluzione Ubar} yields
\begin{equation}
\begin{split}
\bar{U}' =& \phi' \frac{2 f}{2 f + 1} \bigg[
\left(
2 + \frac{(1+z)f'}{2f}
\right) E^2
- \frac{3}{2f} \Omega_{m0} (1+z)^3\\
&- \frac{1}{2f} (1+z) \Omega_{DE}'
\bigg].
\end{split}
\label{movimento}
\end{equation}

Once the function $f(z)$ is specified, Eq.~\eqref{scalarfield} determines $\phi(z)$.  
If the relation can be inverted numerically, one obtains $z=z(\phi)$ and therefore $U(z)$ and $U(\phi)$ from Eq.~\eqref{movimento}. This reconstruction allows one to study the physical viability of the model, in particular the absence of tachyonic instabilities. The effective scalar mass is given by
\begin{equation}
u^2 
\equiv \frac{1}{3}
\left(
\phi \frac{d^2 \bar{U}}{d\phi^2}
     - \frac{d\bar{U}}{d\phi}
\right)
> 0,
\label{scalar mass}
\end{equation}
which follows from Eq.~\eqref{mass} in the specific $f(z)$ framework. 

\medskip

As a diagnostic tool to quantify deviations from the standard $\Lambda$CDM expansion history, the \emph{effective Hubble constant} is defined as
\begin{equation}
\mathcal{H}(z)
= H_0 \, \frac{H_{\text{NEW}}(z)}{H_{\Lambda\text{CDM}}(z)} ,
\label{fit}
\end{equation}
where $H_{\text{NEW}}(z)$ is the expansion rate predicted by the phenomenological $f(z)$ model.  
This ratio captures departures from the $\Lambda$CDM behaviour while keeping the local value of $H_0$ fixed (for a similar approach, see also \citet{kazantzidis2020hints}).  
The apparent redshift dependence of $H_0$ often extracted from Type Ia supernovae comes from the assumption of a $\Lambda$CDM cosmology in the calibration procedure and is therefore not physical. As a consequence it has to be intended as a gap in the standard cosmological model.

\medskip

Finally, two useful cosmographic quantities are introduced and will be used later to constrain some parameters: the present-day deceleration parameter
\begin{equation}
q_0 \equiv -1 + \frac{1}{2}\left. \frac{dE^2}{dz} \right|_{z=0},
\label{deceleration}
\end{equation}
and the jerk parameter
\begin{align}
J_0 & \equiv 1  - \left. \frac{d E^2}{dz} \right|_{z=0} + \frac{1}{2} \left. \frac{d^2 E^2}{dz^2} \right|_{z=0}.
\label{Jerk}
\end{align}

In the $\Lambda$CDM model, these reduce to
\begin{equation}
q_0^{\Lambda\text{CDM}} = -1 + \frac{3}{2}\Omega_{m0},
\qquad
J_0^{\Lambda\text{CDM}} = 1.
\end{equation}

To analyse the model, a phenomenological function $f(z)$ is specified. 
This function enters directly into Eq.~\eqref{eq finale} and, consequently, into Eq.~\eqref{fit}; 
it therefore represents the quantity that can be constrained most directly by the observational data.

When choosing $f(z)$, two physical requirements must be imposed.  
First, as previously explained, the normalisation condition $f(0) = 1$ must be satisfied.  
Second, $f(z)$ should be an increasing function of the redshift, in order to reproduce the observed effective decrease of the Hubble constant $H_{0}$ with redshift, as discussed in Refs.~\citet{dainotti2021hubble, dainotti2022evolution, schiavone2023f, montani2025running}.

The parameters of the model are constrained by fitting the predicted expansion rate 
$\mathcal{H}(z)$ to the binned Pantheon data.  
Since the dataset extends only to relatively low redshift 
(40 bins, with most points below $z \simeq 0.8$ and the highest at $z \simeq 1.2$), 
the function $f(z)$ is modelled as a quadratic Taylor-like expansion,
\begin{equation}
f(z) = 1 + a z + b z^{2},
\label{f(z)}
\end{equation}
which is sufficiently general to capture the behaviour of any function that is differentiable at least twice and has a continuous second derivative near $z=0$.
The coefficients $a$ and $b$ quantify deviations from the standard $\Lambda$CDM scaling, 
while the condition $f(0)=1$ is automatically satisfied. 
A second-order expansion is adopted in order to remain sufficiently general to include both linear effects and possible quadratic corrections. Moreover, from a theoretical point of view, as will be discussed in the following sections, the mass of the scalar degree of freedom depends explicitly on both the first and the second derivatives of the function $f(z)$.
Fixing the expansion at first order would therefore set the second derivative of $f(z)$ identically to zero, restricting the functional freedom of the model and artificially limiting the range of physically admissible scalar-field masses.

\subsection{Model Testing on the Pantheon Sample}
\label{Test1}

The model introduced above is now tested against the binned Pantheon Sample.
In principle, the complete fit would involve five parameters, 
$(a, b, \bar{U}_{0}, \omega_{0}, \omega_{a})$.  Due to the limited constraining capacity of the binned dataset, not all parameters can be freely fitted. To this end, the parameters 
$\omega_{0}$ and $\omega_{a}$ are fixed by requiring that the reconstructed expansion history 
reproduces exactly the $\Lambda$CDM values of the deceleration parameter $q_{0}$ and the jerk 
$J_{0}$. This choice ensures that the model does not deviate too much from the $\Lambda$CDM at $z=0$, and that it behaves well at low redshift. In particular, the model should be able to satisfy the luminosity-redshift relation, according to \citet{riess20162, efstathiou2021h, fazzari2025cosmographic}.
Allowing $\omega_0$ and $\omega_a$ to vary freely, even under Gaussian priors centered on their $\Lambda$CDM values, would result in non-informative posteriors, given the current statistical limitations of the binned supernova sample.\\
Moreover, the parameter $\bar{U}_0 = 0$ is fixed without loss of generality, since the fitting procedure depends only weakly on it. 
Once these cosmographic conditions are imposed, the remaining parameters
$(a, b)$ encode all deviations from standard cosmology.

In previous and related analyses, it has been explicitly verified that applying the same non-linear fitting strategy to the Master Sample yields parameter estimates and associated uncertainties that are fully consistent with those obtained from a Bayesian MCMC approach, with deviations not exceeding the level of $0.07\,\sigma$.
This demonstrates the robustness and mutual consistency of the two statistical methodologies.
While, in principle, an MCMC analysis could also be performed for the Pantheon sample, such an approach would be largely redundant in the present context, since the adopted model is nearly linear in the fitted parameters over the region of interest.
For these reasons, a non-linear fit is employed for the Pantheon sample, in order to ensure consistency with previous analyses performed on the same dataset and to allow a direct and meaningful comparison of the results \citep{dainotti2021hubble, dainotti2022evolution, montani2024metric, montani2025running, montani2025decay}.

\begin{figure}[t]
    \centering
    \includegraphics[width=0.7\textwidth]{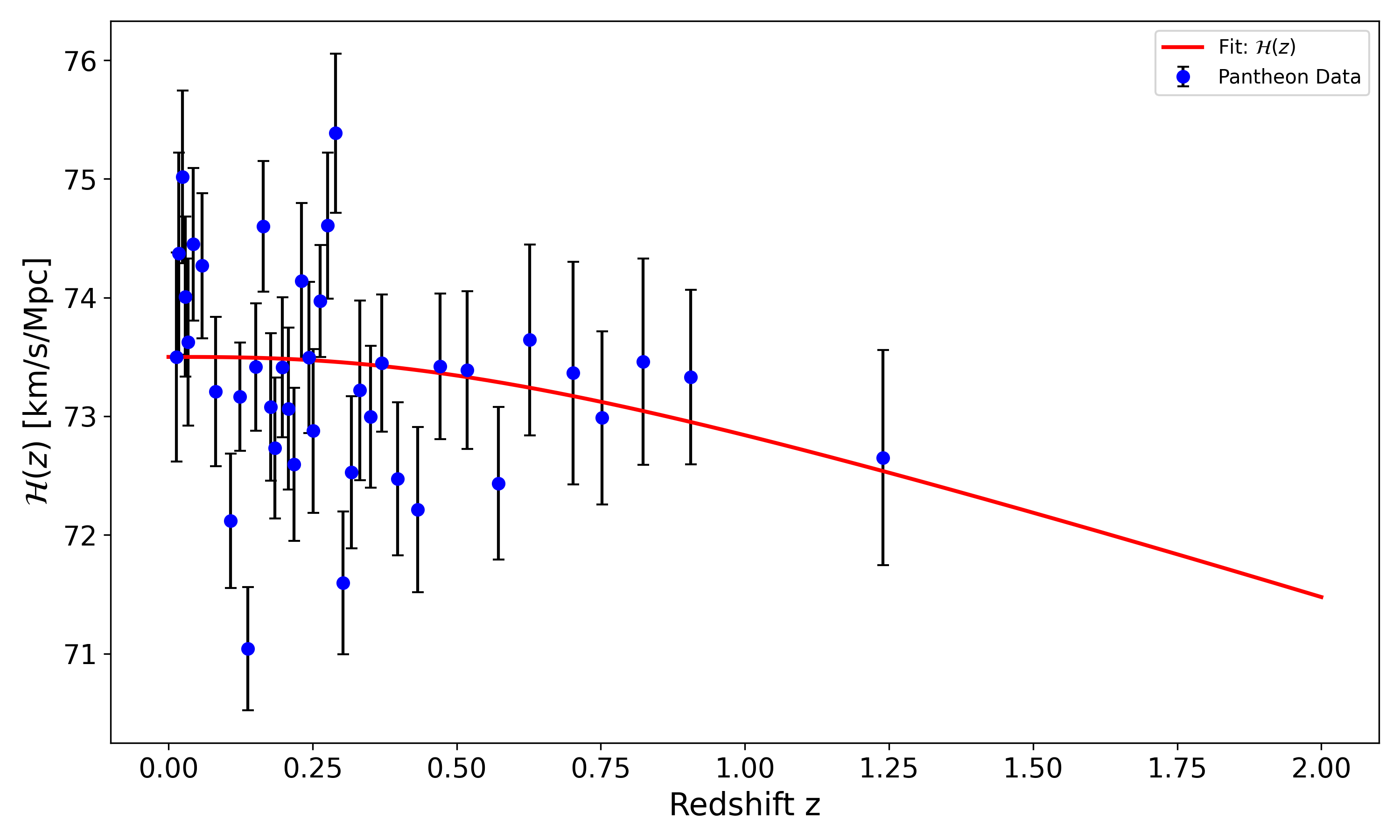}
    \caption{Reconstruction of $\mathcal{H}(z)$ from Tab.~\ref{tab:bestfit} compared with the binned Pantheon data.}
    \label{fig:best_fit}
\end{figure}

The fitting procedure proceeds as follows.  Firstly, a grid is used to sample $b$. 
For every  $ b$, a one parameter non-linear fit determines the fitting value of $a$ by minimising the $\chi^{2}$ computed from the binned Pantheon data.  
Secondly, at each step the associated values of $\omega_{0}$ and $\omega_{a}$ are recomputed from the cosmographic constraints. \\
Finally, after scanning the full grid, all accepted parameter combinations are ranked by their 
$\chi^{2}$ values, and the global minimum is identified.  
The results are summarised in Tab.~\ref{tab:bestfit} and visualised in Fig.~\ref{fig:best_fit}.

\begin{table}[t]
\centering
\begin{tabular}{lc}
\hline
Parameter & Value \\
\hline
$a$        & $0.0212 \pm 0.0068$ \\
$b$        & $0.0033 \pm 0.0022$ \\
$\omega_{0}$    & $-0.989$ \\
$\omega_{a}$    & $0.041$ \\
\hline
\end{tabular}
\caption{Mean values of the parameters of the model with uncertainties. 
The parameters $\omega_0$ and $\omega_a$ do not carry uncertainties because they are derived 
by imposing $J_0=J_0^{\Lambda\mathrm{CDM}}$ and $q_0=q_0^{\Lambda\mathrm{CDM}}$ 
(see Eqs.~\eqref{deceleration}-\eqref{Jerk}).  
The uncertainty associated with $b$ is obtained from the grid resolution, the one associated with $a$ comes from the non-linear fit analysis. As it is can be seen from the table, the parameter $b$ is compatible with zero within two standard deviations.}
\label{tab:bestfit}
\end{table}

\begin{figure}[t]
    \centering
    \includegraphics[width=0.7\textwidth]{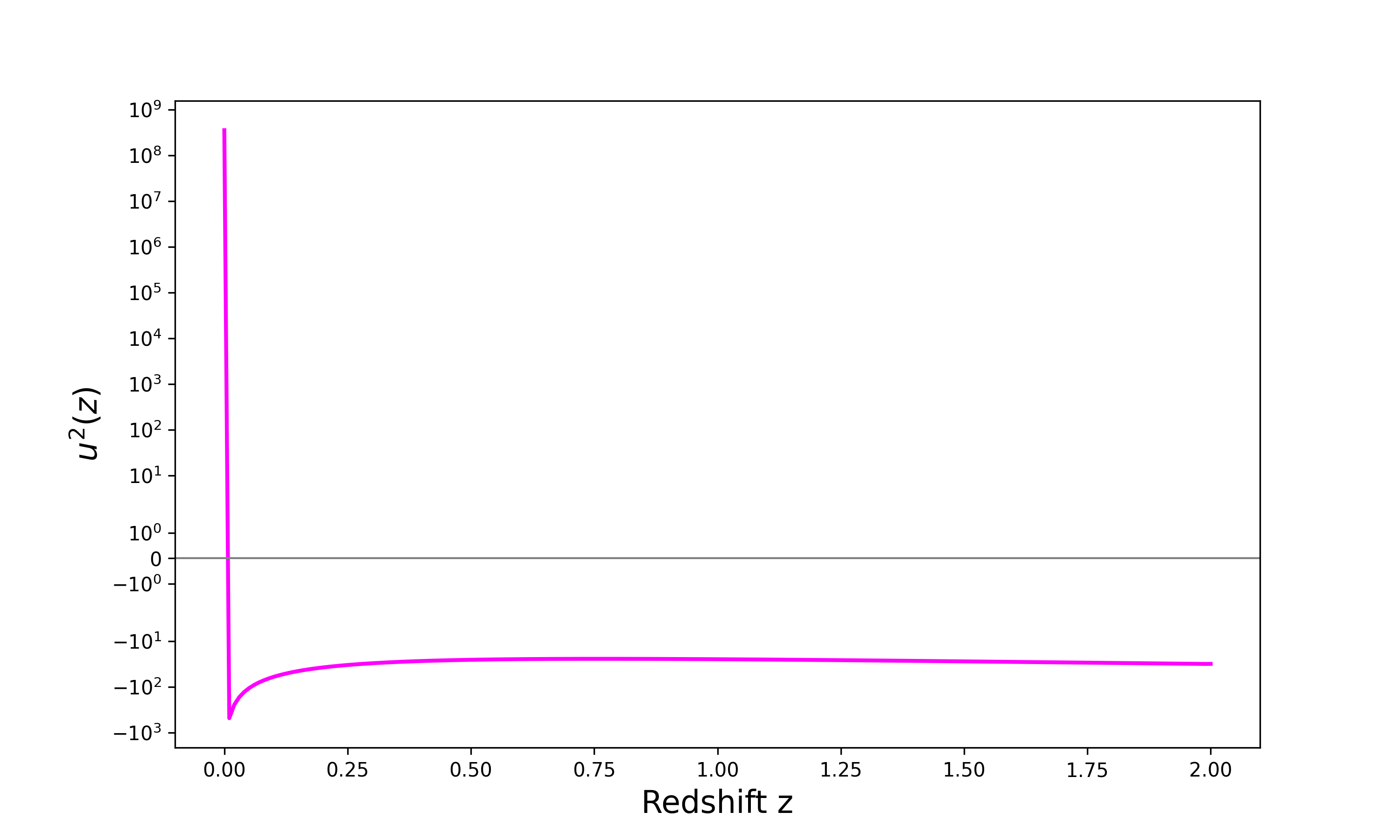}
    \caption{Squared mass of the scalar field for the parameters shown in Fig.~\ref{tab:bestfit}.} 
    \label{fig:best_fit_mass}
\end{figure}

The reduced chi-squared from the data of Tab.~\ref{tab:bestfit} is found to be $\chi^2_{\rm red} = 2.09$. The relatively high value of the reduced $\chi^2$ is explained by the intrinsic scatter of the data.  Similar values have been reported in previous studies, for instance in \citet{montani2025running}, where different cosmological models are compared on the Pantheon sample. Moreover, the $\Lambda$CDM model has $\chi^2_{\,{\rm red},\,\Lambda{\rm CDM}} = 2.17$, hence the model under exam describes the data better than the standard cosmological one.
With these values, the squared mass of the scalar field, defined through 
Eq.~\eqref{scalar mass}, is shown in Fig.~\ref{fig:best_fit_mass}.

To compare the new model with the $\Lambda$CDM, the Akaike (AIC) and Bayesian (BIC) information criteria \citep{akaike2003new,schwarz1978estimating, wagenmakers2007} are defined as 
\begin{equation}
AIC = \chi^2 + 2k,
\qquad
BIC = \chi^2 + k \ln N,
\end{equation}

Tab.~\ref{tab:aic_bic_scale}, based on \citet{Bayes_trotta, jeffreys1998theory}, summarizes the AIC and BIC interval values for which one model is favoured over another.
\begin{table}[h!]
\centering
\caption{Strength of evidence against a model based on $\Delta$AIC and $\Delta$BIC taken from \citet{jeffreys1998theory}.}
\begin{tabular}{c c c}
\hline
$|\Delta \mathrm{AIC}|$ & $|\Delta \mathrm{BIC}|$ & Evidence strength \\
\hline
0 - 2 & 0 - 1       & Inconclusive \\
2 - 4 & 1 - 2.5     & Weak \\
4 - 7 & 2.5 - 5     & Moderate \\
$>$ 7   & $>$ 5       & Strong \\
\hline
\end{tabular}
\label{tab:aic_bic_scale}
\end{table}

The $\Lambda$CDM model gives a reduced chi-squared of 
$\chi^2_{\,{\rm red},\,\Lambda{\rm CDM}} = 2.17$, and hence
\begin{equation}
AIC_{\Lambda CDM}= BIC_{\Lambda CDM}=87.04.
\end{equation}
On the other hand, for the phenomenological $f(R)$ model the values of $AIC$ and $BIC$ are:
\begin{equation}
AIC=85.60,\qquad BIC=87.28.
\end{equation}

According to Tab.~\ref{tab:aic_bic_scale}, the two models are equivalent from a statistical point of view, since
\begin{equation}
\Delta AIC=1.44,\qquad \Delta BIC=-0.24.
\end{equation}
where the convention used here and in the following sections is $\Delta AIC= AIC_{\Lambda CDM}-AIC_{model}$, $\Delta BIC= BIC_{\Lambda CDM}-BIC_{model}$.
From a physical point of view, however, figure~\ref{fig:best_fit_mass} highlights two severe issues:  
(i) the scalar-field mass diverges at $z=0$, and  
(ii) the mass squared becomes negative at $z>0$, signaling a tachyonic instability.  
Both features indicate that the reconstructed model cannot represent a physically acceptable 
metric $f(R)$ theory.  
A deeper investigation is therefore required to understand the origin of these instabilities.

\subsection{Model Testing on the Master Sample}
\label{Test2}

The same analysis is now repeated using the Master Sample, in order to determine whether the unphysical behaviour identified in the Pantheon reconstruction is due to the dataset itself or to an intrinsic limitation of the theoretical framework. \\
The theoretical model remains unchanged (see Eqs.~\eqref{movimento}-\eqref{fit});  
what differs is the statistical methodology and the set of observables used in the likelihood. \\
Specifically, as explained in Sec.~\ref{sec:master_sample} the binning procedure for the Master sample is performed using 20 bins, and both $H_{0}$ and $\Omega_{m0}$ are treated as free parameters and inferred directly from the SNe data divided into bins. This yields 20 values of $H_0$ and 20 values of $\Omega_{m0}$. For this reason, to fit the theoretical $\mathcal{H}(z)$ function for the specific model to the corresponding binned data samples, the likelihood is constructed in order to
minimize the total chi-squared 
    \begin{equation}    \chi^2_{tot}=\chi^2_{H_{0_{bin}}}+\chi^2_{\Omega_{m^0_{bin}}} \,,
    \end{equation}
where $X_{\text{bin}}$ represents the single data value in each bin, i.e., the binned values of $H_0$ and $\Omega_{m}^0$. This is the same procedure adopted in \cite{fazzari2025effective}.
This implies that when we compute the reduced chi-squared $\chi^2_{red}=\chi^2/\text{d.o.f.}$, the number of degree of freedom is $d.o.f.=N_{data} - N_{par}$ but $N_{data}=2N_{bins}$, since the dataset consists of 20 values of $H_0$ and 20 values of $\Omega_{m0}$, which are independent of each other \citep{verde2010statistical}.


The MCMC analysis is performed using the \texttt{Cobaya} tool, which integrates the background 
equations of the model at each likelihood evaluation and updates the parameter vector through the 
sampling algorithm.  
Multiple chains are evolved in parallel until convergence is obtained, which is monitored through the Gelman-Rubin statistic \citep{gelman1992inference}. Convergence is reached when the potential scale reduction factor satisfies $R - 1 < 0.01$.

\begin{figure}[t]
    \centering
    \includegraphics[width=0.7\textwidth]{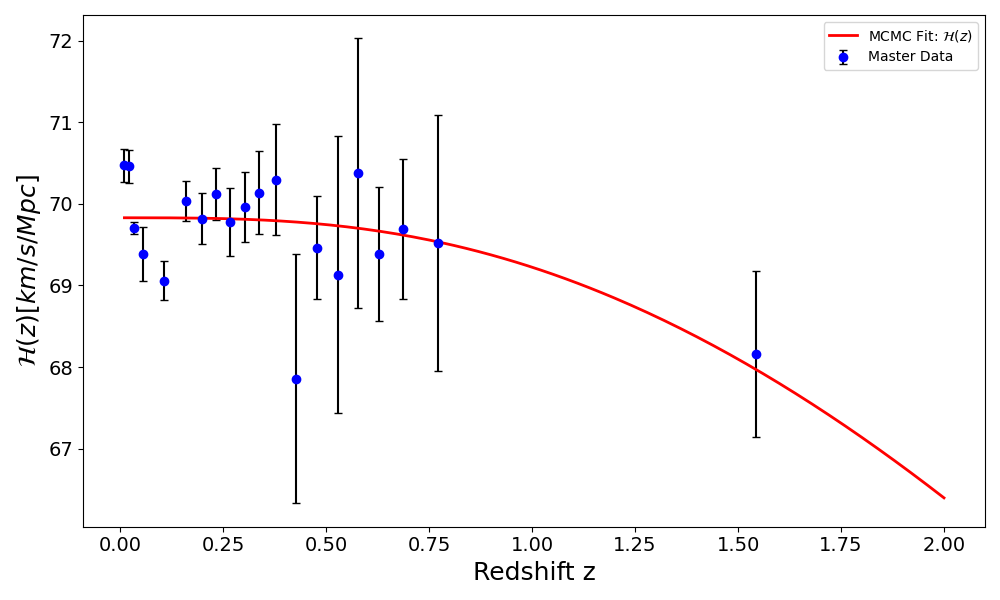}
    \caption{Reconstruction of $\mathcal{H}(z)$ from Tab.~\ref{tab:master_bestfit} compared with the binned Pantheon data.}
    \label{fig:master_Hz}
\end{figure}

As in the previous section, an external grid in $ b$ is constructed due to the limited constraining power of the data. Being a second-order term in the expansion of $f(z)$, $b$ has a smaller impact near low redshift than the linear coefficient; therefore, it is selected for the grid, while the dominant linear coefficient $a$ is allowed to vary freely in the MCMC.  
For each point, a full MCMC scan of the remaining 
parameter space $(H_{0},\Omega_{m0},a)$ is carried out. For $H_0$ a flat prior between $[60,80]$ is used, as well as for the parameter $a$ in the range $[-1,1]$. For the parameter $\Omega_{m0}$, a Gaussian prior $\Omega_{m0}=0.322 \pm 0.025$ ($5\sigma$) is assumed, to be consistent with the data analysis carried out to obtain the binned Master Sample.\\
For each trial configuration $(a, b, H_{0}, \Omega_{m0})$, the values of 
$\omega_{0}$ and $\omega_{a}$ are computed analytically by imposing that the deceleration and jerk 
parameters match those of the $\Lambda$CDM at $z=0$
(Eqs.~\eqref{deceleration}-\eqref{Jerk}).  Again, $\bar{U}_0=0$ is fixed.
After integrating the background dynamics, obtaining $H(z)$ and evaluating the $\chi^{2}$ of 
$\mathcal{H}(z)$ on the Master Sample, the best posterior values are recorded.  
The optimal five-parameter configuration is then selected as the one minimising the mean $\chi^2$.
The reconstruction of $\mathcal{H}(z)$ obtained from the Master Sample in this case is shown in 
Fig.~\ref{fig:master_Hz}.

The corresponding parameter values are summarised in Tab.~\ref{tab:master_bestfit}.

\begin{table}[t]
\centering
\begin{tabular}{lc}
\hline
Parameter & Value \\
\hline
$H_{0}$ [km\,s$^{-1}$\,Mpc $^{-1}$]        & $69.830 \pm 0.060$ \\
$\Omega_{m0}$   & $0.3185 \pm 0.0048$ \\
$a$             & $-0.025 \pm 0.019$ \\
$b$             & $0.0360 \pm 0.0060$ \\
$w_{0}$         & $-1.0122 \pm 0.0093$ \\
$w_{a}$         & $-0.014 \pm 0.037$ \\
\hline
\end{tabular}
\caption{Posterior mean values and associated uncertainties of the model parameters obtained from the 
MCMC analysis of the Master Sample.}
\label{tab:master_bestfit}
\end{table}

\begin{figure}[t]
    \centering
    \includegraphics[width=0.7\textwidth]{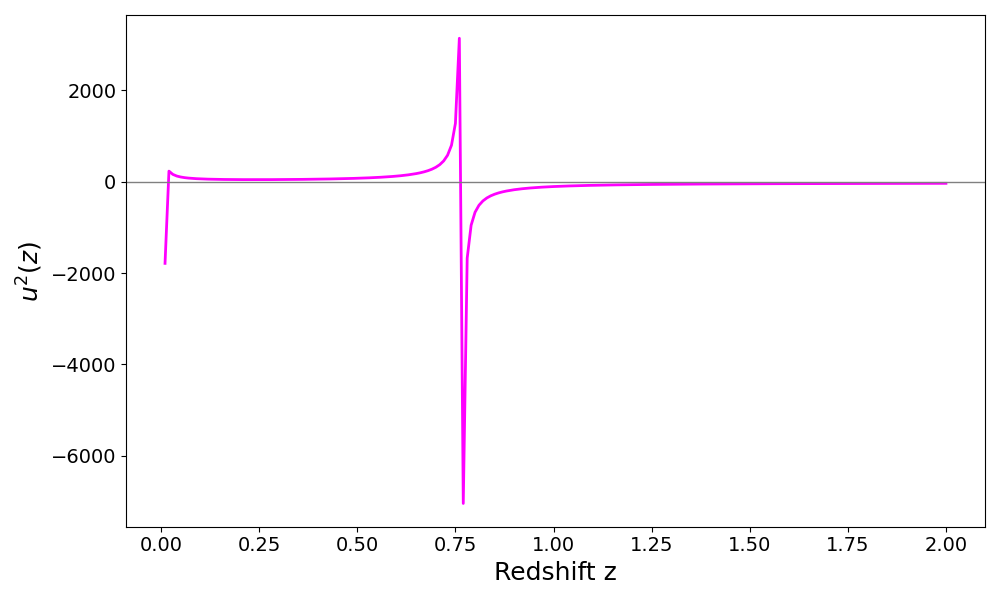}
    \caption{Scalar field mass squared $u^{2}(z)$ obtained from the MCMC analysis of the Master Sample.}
    \label{fig:master_mass}
\end{figure}
  
Regarding the statistical analysis, the reduced chi-squared of the best-fit case is
\begin{equation}
\chi^{2}_{\rm red} = 1.20,
\end{equation}
which leads to:
\begin{equation}
AIC=50.04,\qquad BIC=53.39.
\end{equation}
On the contrary, the values obtained for the $\Lambda$CDM 
are $\chi^2_{\rm red,\,\Lambda CDM} = 2.28$ and
\begin{equation}
AIC_{\Lambda CDM}=47.60,\qquad BIC_{\Lambda CDM}=48.59.
\end{equation}
Hence
\begin{equation}
\Delta AIC=-2.44,\qquad \Delta BIC=-4.80.
\end{equation}
Indicating a weak preference for the AIC and a moderate one for the BIC in favor of the $\Lambda$CDM, according to Tab.~\ref{tab:aic_bic_scale}.

The scalar field mass squared reconstructed from the MCMC chains is displayed in 
Fig.~\ref{fig:master_mass}.
Again, as in the Pantheon case, the reconstructed scalar field mass exhibits both  
(i) a divergence at $z=0$, and  
(ii) a tachyonic regime where $u^{2}(z) < 0$ for $z>0$.  
The two datasets therefore lead to the same conclusion:  
the phenomenological model based on $f(z)$ cannot represent a viable metric $f(R)$ theory.

A sharp peak in $u^2(z)$ is observed around $z \sim 0.75$. 
This feature originates from the structure of the reconstruction equations: 
the effective mass $u^2$ depends on both the first and second derivatives of the scalar field potential $\bar{U}(\phi)$, 
which are functions of $\phi(z)$ and the phenomenological function $f(z)$. 
Near this redshift, the combination of small denominators in Eq.~\eqref{eq:tosta} and the subtraction in Eq.~\eqref{scalar mass} 
leads to a transient numerical amplification.

\subsection*{Joint conclusions from Pantheon and Master analyses}

The analyses based on both the Pantheon and the Master samples indicate that the model introduced in Sec.~\ref{model descr1} is statistically capable of reproducing the observed behaviour of the effective $H_0$ at low redshift, providing a good fit.  
However, the reconstruction of the scalar field and its potential reveals two severe issues concerning the physical viability of the model.  
These problems arise intrinsically from the structure of the theory and appear independently of the adopted fitting procedure and of the dataset, whether non-linear regression (Pantheon sample) or MCMC sampling (Master sample).

In both analyses, the reconstructed squared mass becomes negative over extended redshift intervals.  
This behaviour is a direct consequence of requiring the effective dark energy parameters to reproduce the $\Lambda$CDM at $z=0$, namely
\begin{equation}
q_{0}=q_{0}^{\Lambda{\rm CDM}}, \qquad 
J_{0}=J_{0}^{\Lambda{\rm CDM}},
\label{eq:LCDM_constraints}
\end{equation}
which imposes strong constraints on the derivatives of the reconstructed potential.  
Relaxing these conditions would in principle avoid tachyonic instabilities, but only at the cost of departing from the $\Lambda$CDM already at $z=0$, which would defeat the purpose of modelling a controlled deviation from the standard scenario.

A second, equally severe issue is the systematic divergence of the scalar mass as $z \rightarrow 0$.  
To clarify the origin of this divergence, Eq.~\eqref{scalar mass} must be examined in detail.  
The most relevant contribution comes from the first term,
\begin{align}
\frac{d^2 \bar U}{d\phi^2}
= \frac{1}{\frac{d\phi}{dz}}
\Bigg\{
& \Bigg[
\left( \frac{1}{2} - \frac{4f}{2f+1} \right)\frac{df}{dz}
- (1+z)\frac{1}{2f+1}\left( \frac{df}{dz} \right)^2
\nonumber \\[2pt]
&  + \frac{1+z}{2}\frac{d^2 f}{dz^2}
\Bigg] 
\frac{2E^2}{2f+1}
\nonumber \\[2pt]
& + \Omega_{m0}(1+z)^3 \frac{2}{(2f+1)^2}\frac{df}{dz} 
\nonumber \\[2pt]
& + \frac{3\Omega_{m0}(1+z)^3}{(2f+1)f}\frac{df}{dz} 
+ \frac{1+z}{(2f+1)f} \frac{df}{dz}\frac{d\bar U}{dz} 
\nonumber \\[2pt]
& + \frac{4}{f(2f+1)} \frac{d\bar U}{dz} 
+ \frac{3\Omega_{m0}(1+z)^2}{2f+1}
\Bigg\},
\label{eq:tosta}
\end{align}
while the second term takes the simpler form
\begin{equation}
\frac{d\bar{U}}{d\phi}
= \frac{2f}{2f+1}
\left[
\left( 2 + \frac{(1+z)}{2f} \frac{df}{dz} \right) E^2
- \frac{3}{2f}\Omega_{m0} (1+z)^3
\right].
\label{eq meno tosta}
\end{equation}

It follows from Eq.~\eqref{eq:tosta} that the second derivative of the potential with respect to the scalar field, and therefore the mass of the scalar degree of freedom, depends explicitly on the second derivative of the function $f(z)$. This provides the theoretical motivation for adopting a quadratic expansion of $f(z)$, as introduced in Eq.~\eqref{f(z)}, since a purely linear parametrisation would fix this contribution identically to zero.
It is evident that Eq.~\eqref{eq:tosta} diverges as $z \rightarrow 0$, because $\frac{d\phi}{dz} \rightarrow 0^{-}$ due to the initial conditions imposed in Eq.~\eqref{problemacauchy} and due to the fact that $f(z)>0$ (see Eq.~\eqref{eq finale}) ensures $\phi'<0$. 
Therefore, for sufficiently small redshift values this term dominates the behaviour of $u^2$.

Since the sign of the scalar mass is determined by $\frac{d^2\bar U}{d\phi^2}$ and $\frac{d\phi}{dz}<0$, avoiding a tachyonic instability requires
\begin{align}
Q \equiv \Bigg\{ &
\Big[ 
\left( \tfrac12 - \tfrac{4f}{2f + 1} \right) \tfrac{df}{dz} 
- (1 + z)\tfrac{1}{2f + 1}\left( \tfrac{df}{dz} \right)^2 \\[2pt]
&+ \tfrac{1 + z}{2} \tfrac{d^2 f}{dz^2} 
\Big] 
\frac{2E^2}{2f + 1} 
\nonumber \\[2pt]
&+ \Omega_{m0} (1 + z)^3 \frac{2}{(2f + 1)^2} \frac{df}{dz} \\[2pt]
&+ \frac{3 \Omega_{m0} (1 + z)^3}{(2f + 1)f} \frac{df}{dz} 
\nonumber \\[2pt]
& + \frac{(1 + z)}{(2f + 1)f} \frac{df}{dz} \frac{d\bar{U}}{dz} 
+ \frac{4}{f(2f + 1)} \frac{d\bar{U}}{dz} \\[2pt]
&+ \frac{3 \Omega_{m0} (1 + z)^2}{2f + 1}
\Bigg\} < 0 .
\label{eq: disequazione}
\end{align}

Although a function $f(z)$ satisfying all physical requirements (e.g.\ $f(0)=1$) and the inequality~\eqref{eq: disequazione}) may exist in principle, it must also reproduce the observational constraints derived from Pantheon through Eq.~\eqref{movimento}.  
This compatibility condition is highly restrictive.

\begin{figure}[t]
\centering
\includegraphics[width=0.7\textwidth]{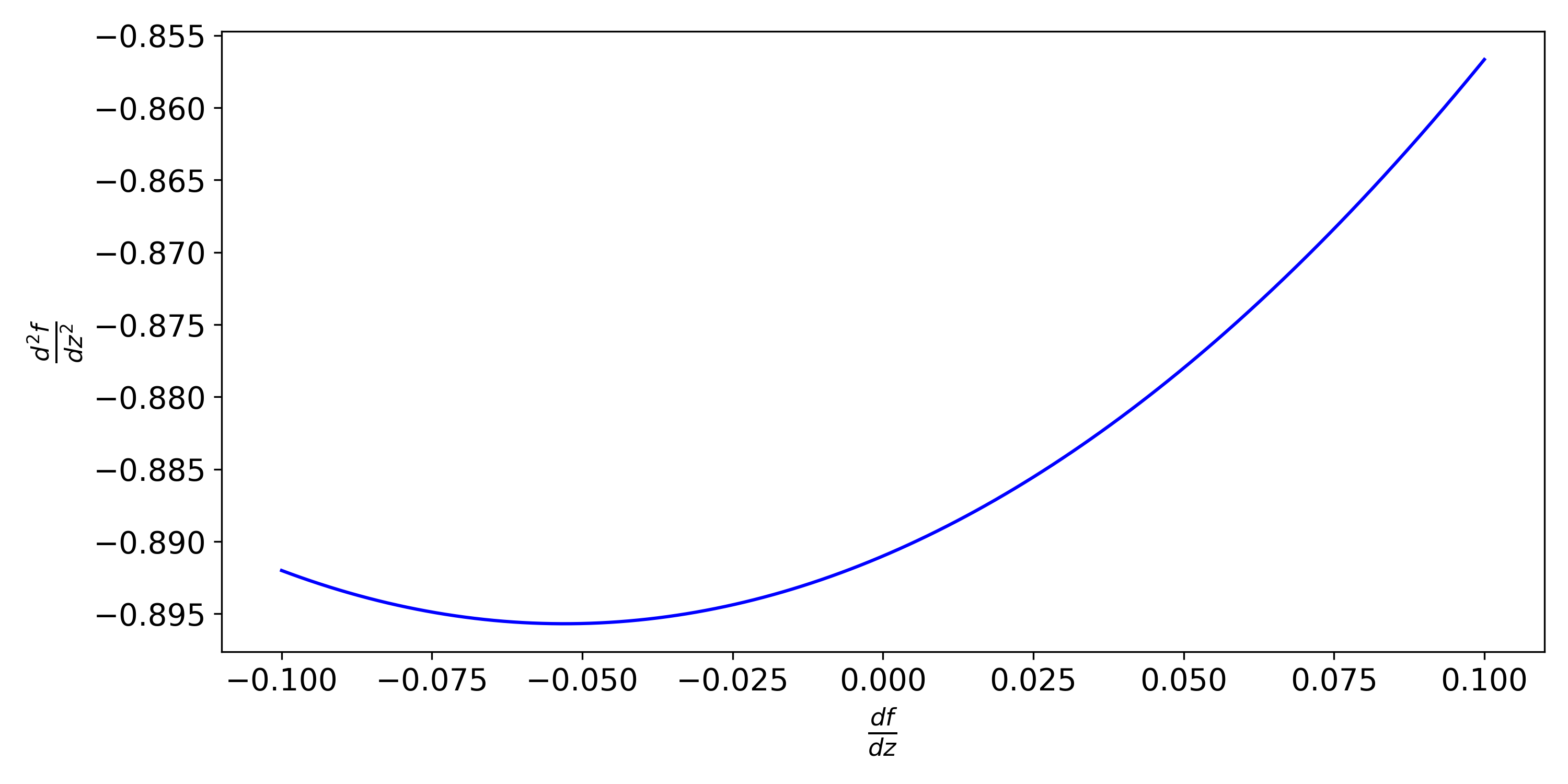}
\caption{Solution of Eq.~\eqref{eq:massa_finitudine} for $\alpha=1$, expressing the required relation between $\frac{df}{dz}$ and $\frac{d^{2}f}{dz^{2}}$ at $z=0$ for a finite scalar mass.  
From the initial conditions of the problem, $\phi'(0)=0$, see Eq.~\eqref{problemacauchy}.}
\label{fig: seconda in funzione di prima}
\end{figure}

Furthermore, ensuring that the scalar mass remains finite today requires solving
\begin{align}
Q|_{z=0} = \alpha\,\frac{d\phi}{dz}\sim\mathcal{O}(1)\cdot\frac{d\phi}{dz},
\label{eq:massa_finitudine}
\end{align}
which can be solved algebraically for $\frac{d^2 f}{dz^2}$ as a function of $\frac{df}{dz}$.  
The resulting relation, shown in Fig.~\ref{fig: seconda in funzione di prima}, specifies the allowed combinations of derivatives at $z=0$ for a finite scalar mass (for a chosen value of $\alpha$).
This figure shows that, for sufficiently small values of $\frac{df}{dz}$, the second derivative must satisfy $\frac{d^{2}f}{dz^{2}} < 0$ to prevent a divergence.  
However, the mean values of the parameters extracted from both Pantheon~(\ref{tab:bestfit}) and the Master sample~(\ref{tab:master_bestfit}) yield values of $\frac{df}{dz}$ incompatible with any $\frac{d^{2}f}{dz^{2}}$ satisfying Eq.~\eqref{eq:massa_finitudine}.

Therefore, the conclusion is robust: the model cannot simultaneously satisfy  
(i) proximity to the $\Lambda$CDM at $z=0$,  
(ii) positivity of the squared scalar mass, and  
(iii) finiteness of the mass at $z=0$.  
In particular, while conditions (i) and (ii) may be traded against each other, requirement (iii) cannot be fulfilled because the behaviour $\frac{d\phi}{dz}|_{z=0}=0$ (see Eq.~\eqref{problemacauchy}) forces Eq.~\eqref{eq:tosta} to diverge.

The analyses performed on both the Pantheon and the Master samples show that the model introduced in Sec.~\ref{model descr1} is statistically able to reproduce the observed behaviour of the effective $H_0$ at low redshift.  
However, the reconstruction of the scalar field and its potential reveals two severe physical viability issues that appear robustly in both datasets.  
This approach to $f(R)$ theories should hence be discarded, and a new formalism will be presented in the following sections.

It is important to note that even if, instead of imposing $\phi(z=0)=1$, one takes $\phi(z=0)=1\pm10^{-7}$(see \citet{hu2007models}), the problems described in this section still remain. The $10^{-7}$ deviation would not be sufficient to avoid the unphysical value of the mass at $z=0$, since the value of the derivative of the scalar field would be $\phi'(z=0)=\pm10^{-7}$.

\section{A Proposal for a Viable Model}
\label{sec:newmodel}

In Sec.~\ref{phenomenology}, $f(R)$ theories were introduced by choosing a function $f(z)$ from which the scalar field, its potential, and finally the function $f(R)$ were reconstructed.  
The analysis suggested the presence of a no-go theorem: in this framework, modified-gravity models of this type are unable to reproduce the observational data without developing severe physical inconsistencies, most notably a pathological behaviour of the scalar field mass.  
A fundamental tension therefore emerged: either the observational fit is accurate while the theory becomes physically unviable, or the requirement of a finite and positive mass is imposed, at the price of a substantial deviation from the data.

It is important to stress that the procedure adopted in Sec.~\ref{phenomenology}, although fairly general, is not the most universal approach to constructing $f(R)$ theories.  
The mapping between a function $f(z)$ and a function $f(R)$ requires a specific reconstruction algorithm that necessarily introduces choices, including the selection of integration constants.  
Once these are fixed, the theory becomes dependent on the adopted prescriptions.

It is evident that the main difficulty of the previous formalism comes from the highly restrictive initial conditions imposed on Eq.~\eqref{problemacauchy}, which leave no freedom in the initial value of the scalar field derivative at $z=0$.  
Enforcing $\phi'(z=0)=0\pm10^{-7}$ inevitably drives the effective mass of the field to diverge at the present epoch, rendering the model physically inconsistent.  
Since the scalar field dynamics are governed by a second-order differential equation, as follows from Eq.~\eqref{eomphi2}, fixing $\phi'$ \emph{a priori} removes too many degrees of freedom and overconstrains the Cauchy problem.

Relaxing the functional constraint on $f(z)$ is not possible, as this would give problems to the reconstructed Hubble function $\mathcal{H}(z)$.  
On the other hand, releasing the condition on $\phi(0)$ would cause a great deviation from the GR even at low redshift. 

The only robust solution is therefore to construct a new framework that removes the function $f(z)$ from the formulation entirely.  
A direct consequence of this choice is a mild departure from General Relativity at $z=0$, since the Einsteinian limit, characterized by $\phi(z=0)=1$ and $\phi'(z=0)=0$, cannot be fully recovered.

The condition $\phi(z=0)=1$ still holds true, while $\phi'(z=0)=0$ can now be consistently discarded.  
Such a departure from General Relativity at the present cosmological epoch does not constitute a critical issue.  
The theory proposed here is to be interpreted as an effective description, valid only within a restricted domain both in space (on cosmological scales) and in time (across the relevant redshift interval).  
General Relativity is recovered on local scales, such as within the Solar System, due to the efficiency of the chameleon mechanism, which renders the scalar field extremely massive in high-density environments and suppresses any deviations from Einsteinian gravity.

For these reasons, an alternative formulation of modified gravity is adopted in the present chapter, following the approach introduced in~\citet{montani2025decay}.

\subsection{Model Description}

Starting from Eq.~\eqref{friedmann}, the system is now closed by introducing the dynamical condition
\begin{equation}
6H\dot{\phi}=U(\phi),
\label{eq:condition_new}
\end{equation}
motivated by its ability to bring the Friedmann equation close to the standard GR form, with the scalar field acting only as a rescaling factor that accounts for the effective behaviour of $H_0$. With this condition, $U(\phi(z))$ becomes a dynamical quantity determined a posteriori once $\phi(z)$ is known.  
Under the condition of Eq.~\eqref{eq:condition_new}, the Friedmann equation reduces to
\begin{equation}
H^{2}= \frac{\kappa}{3\phi}\,\rho_{\rm tot},
\label{eq:Friedmann_reduced}
\end{equation}
where $\rho_{\rm tot}$ is the total energy density contribution.
The CPL parametrization (Eq.~\eqref{EOS}) is adopted for the dark energy sector.  
The continuity equation integrates to Eq.~\eqref{OmegaDE}, while matter evolves as usual,
\begin{equation}
    \rho_m(z)=\rho_{m,0}(1+z)^3.
\end{equation}
Thus,
\begin{equation}
\rho_{\rm tot}(z)\equiv\rho_m(z)+\rho_{\rm DE}(z).
\end{equation}
The potential satisfies the differential equation
\begin{equation}
\bar U'=\left(-\frac{2}{1+z}+(\ln E)' \right)\bar U,
\end{equation}
whose solution is
\begin{equation}
\bar U(z)=-\gamma\,\frac{E(z)}{(1+z)^2}, 
\qquad \gamma>0.
\end{equation}

The scalar field evolves according to
\begin{equation}
\frac{d\phi}{dz}= \frac{\gamma}{E(1+z)^2},
\qquad \phi(0)=1,
\label{eq:phi_evol}
\end{equation}
together with
\begin{equation}
E^2=\frac{\Omega_{\rm tot}}{\phi(z)},
\label{equadro}
\end{equation}
where $\Omega_{\rm tot}=\Omega_{\rm DE}+\Omega_m$.
The condition $\gamma>0$ is imposed because Eq.~\eqref{equadro} should be a decreasing function of the redshift. Hence, $\phi$ should increase and then the condition follows from Eq.~\eqref{eq:phi_evol}.
Again, $\Omega_m\equiv\Omega_{m0}(1+z)^3$, $\Omega_{m}^0 \equiv \kappa \rho_{m0} / (3 H_0^2)$, $\Omega_{DE}\equiv \kappa \rho_{DE} / (3 H_0^2)$.\\
To avoid an excessive deviation from the $\Lambda$CDM at $z=0$ and to prevent introducing too many free parameters given the constraining power of the data, $\omega_0$ and $\omega_a$ are fixed by matching the deceleration and jerk parameters to their values in the standard cosmological model at $z=0$, following the same procedure described in Sec.~\ref{phenomenology}.

From the deceleration parameter at $z=0$ one obtains
\begin{equation}
\omega_0=\frac{\gamma}{3(1-\Omega_{m0})}-1,
\label{eq:w0_fix_new}
\end{equation}
while the jerk condition yields
\begin{equation}
\omega_a = 
3\omega_0\left(\frac{\gamma}{2}-1\right)
 - 3\omega_0^2
 - \frac{\gamma}{6}\frac{1+3\gamma}{1-\Omega_{m0}}.
\label{eq:wa_fix_new}
\end{equation}

The Hubble function is reconstructed through Eq.~\eqref{fit}.

\subsection{Model Testing on the Pantheon Sample}
\label{Test3}
Similarly to Sec.~\ref{Test1}, the model is tested on the Pantheon Sample using a nonlinear fitting procedure. $\Omega_{m0}=0.298$ is again fixed for the Pantheon Sample.\\
As explained in Sec.~\ref{Test1}, the statistical analysis is performed with a non-linear fit.\\

It is worth stressing that the numerical exploration is effectively performed on the parameter $\gamma$. 
The reason is twofold. 
First, the parameters $\omega_0$ and $\omega_a$ are \emph{de facto} fixed by requiring that the present-day values of the cosmographic parameters $q_0$ and $j_0$ lie within $10\%$ of their $\Lambda$CDM values. 
Second, the parameter $\gamma$ is directly related, through Eq.~\ref{eq:phi_evol}, to the derivative of the scalar field $\phi'$, and therefore controls the new physical features introduced by the additional dynamical condition~(\ref{eq:condition_new}). 

A grid of $(\omega_0,\omega_a)$ values, centered around the predictions of Eqs.~\eqref{eq:w0_fix_new}–~\eqref{eq:wa_fix_new}, is constructed and allowed to vary by up to $10\%$.
For each pair $(\omega_0,\omega_a)$, the parameter $\gamma$ is determined by a nonlinear least squares minimization of $\chi^2$.

A model is accepted only if the effective scalar mass condition (Eq.~\eqref{scalar mass}) is fulfilled.
Among all accepted configurations, the parameter set with the minimum $\chi^2$ is selected.

\begin{figure}[htbp]
    \centering
    \includegraphics[width=0.7\textwidth]{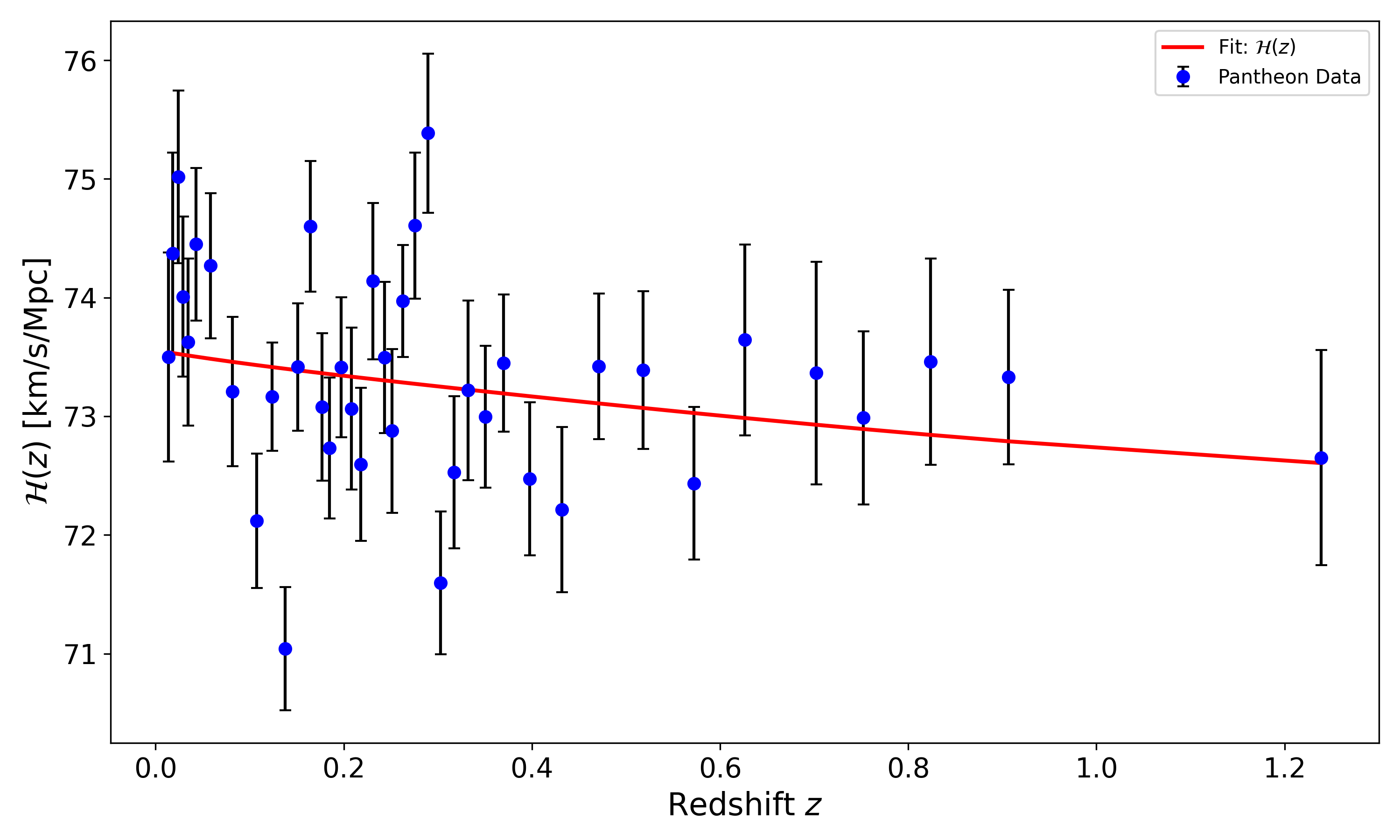}
    \caption{Reconstruction of $\mathcal{H}(z)$ from Tab.~\ref{tab:CPL_bestfit} compared with the binned Pantheon data.}
    \label{fig:Hubble_fit3}
\end{figure}

\begin{figure}[htbp]
    \centering
    \includegraphics[width=0.7\textwidth]{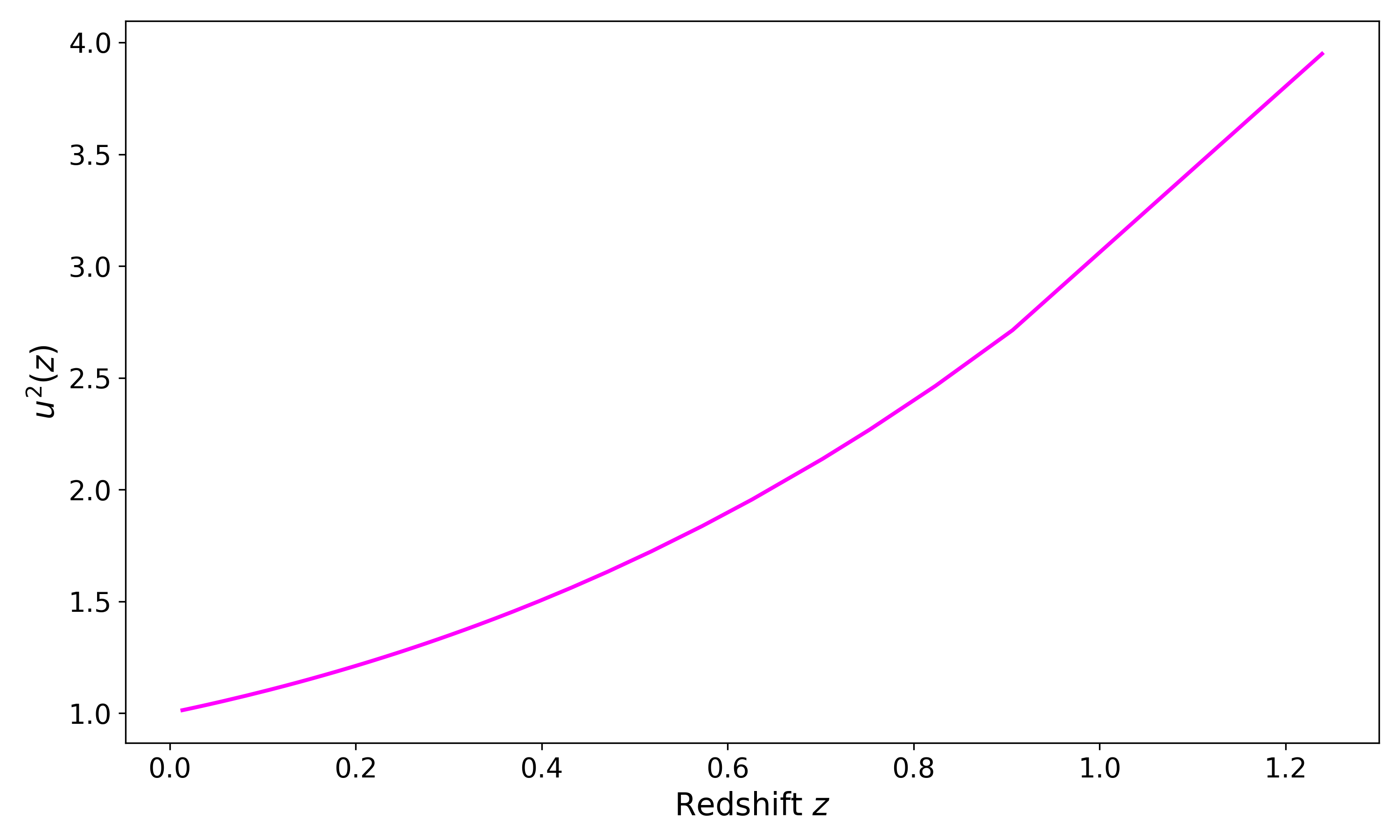}
    \caption{Squared mass of the scalar field for this model.}
    \label{fig:mass_plot3}
\end{figure}

The resulting mean values of the fit parameters are reported in Tab.~\ref{tab:CPL_bestfit}, while the same results are shown in Fig.~\ref{fig:Hubble_fit3}.

\begin{table}[t]
\centering
\begin{tabular}{lc}
\hline
Parameter & Value \\
\hline
$\gamma$        & $0.114 \pm 0.017$ \\
$\omega_{0}$    & $-0.963 \pm 0.020$ \\
$\omega_{a}$    & $-0.034 \pm 0.020$ \\
\hline
\end{tabular}
\caption{Mean values of the parameters of the model with uncertainties. The uncertainty on $\gamma$ arises from the non-linear fit results, while those on the others parameter come from the grid width. The uncertainty on the parameter $\gamma$ comes from the non-linear fit analysis, while those on the other parameters are fixed by the grid width.}
\label{tab:CPL_bestfit}
\end{table}

The corresponding cosmographic parameters read 

\begin{equation}
q_0 = -0.571 \pm 0.021,
\qquad
J_0 = 1.064 \pm 0.083.
\end{equation}

The reduced chi-square (from the parameters of Tab.~\ref{tab:CPL_bestfit}) is
\begin{equation}
    \chi^2_{\rm red}=2.00.
\end{equation}

Again, the value of the $\chi^2_{red}$ is high due to the intrinsic scatter of the data, but it is lower than the $\chi^2_{\,{\rm red},\,\Lambda{\rm CDM}} = 2.17$.

The scalar field mass in Fig.~\ref{fig:mass_plot3} remains positive and finite for all redshifts, with a magnitude consistently of order $\mathcal{O}(1)$.
Therefore, the model is physically viable.

The cosmographic quantities deviate by less than one standard deviation from the $\Lambda$CDM predictions, showing that the scenario remains close to the standard cosmological model while allowing for an effective running of $H_{0}$.

For this model,
\begin{equation}
AIC=80.08,\qquad BIC=81.77.
\end{equation}

For the $\Lambda$CDM on the same dataset,
\begin{equation}
AIC_{\Lambda CDM}=BIC_{\Lambda CDM}=87.04.
\end{equation}

The differences,
\begin{equation}
\Delta AIC \simeq 7, \qquad \Delta BIC \simeq 5.3,
\end{equation}
constitute strong positive statistical evidence in favour of the modified gravity model.

\subsection{Model Testing on the Master Sample}
\label{Test4}

Given the excellent agreement obtained with the Pantheon dataset, the model is now tested on the Master Sample using a MCMC fitting procedure with \textsc{Cobaya}, consistently with the approach taken for the first model analysed in Sec.~\ref{Test2}.
The reconstruction again uses the system of Eqs.~\eqref{fit},~\eqref{equadro},~\eqref{eq:phi_evol}.\\
Regarding prior ranges of the free parameters we adopted for $H_0$ a flat prior between $[60,80]$, for $\gamma$ a uniform prior $(0,1]$, while a Gaussian prior $\Omega_{m0}=0.322 \pm 0.025$ ($5\sigma$) is assumed. Again, this choice of priors was made in order to remain consistent with those used in the construction of the Master sample.

\begin{figure}[htbp]
    \centering
    \includegraphics[width=0.7\textwidth]{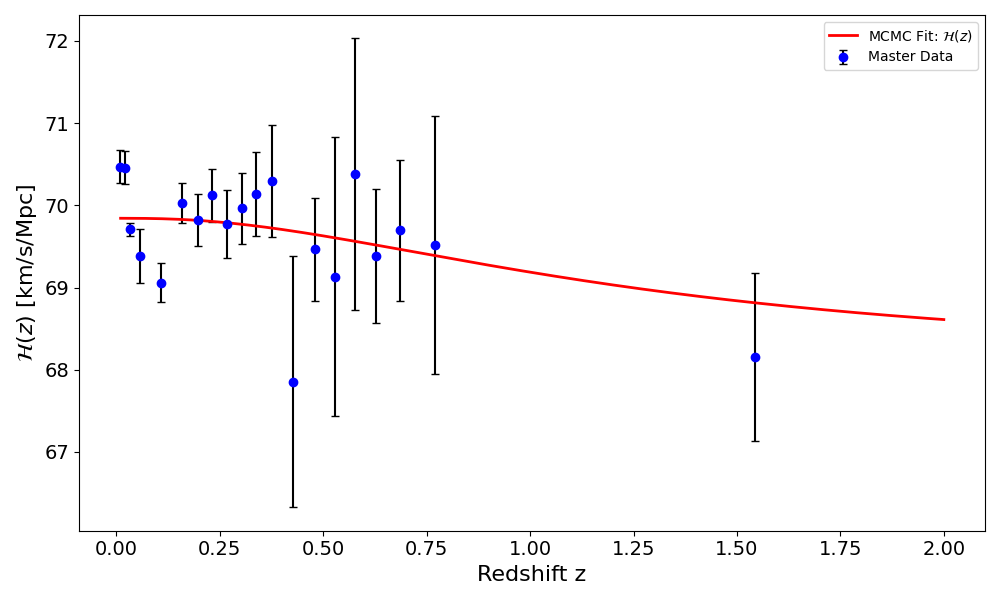}
    \caption{Reconstruction of $\mathcal{H}(z)$ from Tab.~\ref{tab:master_bestfit_all} compared with the binned Master data sample.}
    \label{fig:masterbestgrafico}
\end{figure}

The mean values of the fit parameters values for the model parameters are collected in Table~\ref{tab:master_bestfit_all}, and shown in Fig.~\ref{fig:masterbestgrafico}.

\begin{table}[t]
\centering
\begin{tabular}{lc}
\hline
Parameter & Value \\
\hline
$H_{0}$ [km\,s$^{-1}$\,Mpc $^{-1}$]           & $69.790 \pm 0.060$ \\
$\Omega_{m0}$     & $0.3186 \pm 0.0049$ \\
$\gamma$          & $0.14 \pm 0.10$ \\
$\omega_{0}$      & $-0.932 \pm 0.049$ \\
$\omega_{a}$      & $-0.053 \pm 0.005$ \\
\hline
\end{tabular}
\caption{Mean values with associated uncertainties obtained from the MCMC analysis of the model on the Master Sample.}
\label{tab:master_bestfit_all}
\end{table}

The parameters $\omega_0$ and $\omega_a$ remain compatible with the $\Lambda$CDM predictions at approximately the $1\sigma$ level.

\begin{figure}[htbp]
    \centering
    \includegraphics[width=0.7\textwidth]{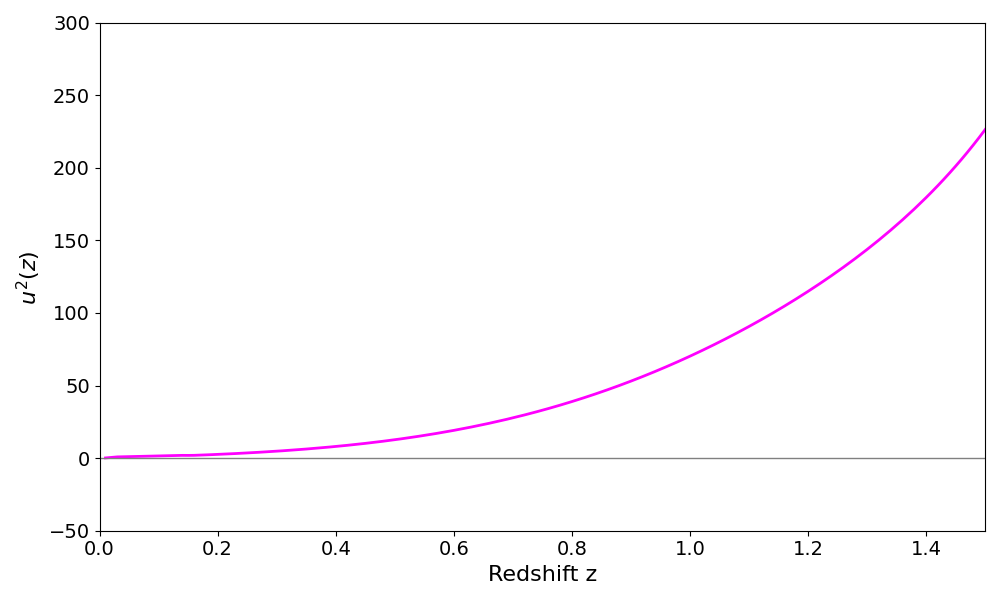}
    \caption{Squared mass of the scalar field for the Master Sample fit.}
    \label{fig:masterbestgraficomassa}
\end{figure}

The reduced chi-square from the fit is found to be
\begin{equation}
\chi^2_{\rm red}=1.11 \qquad (\chi^2=41.09),
\end{equation}

Moreover, the model remains physically consistent: the scalar field mass stays positive and monotonic, never diverging throughout the full redshift interval (see Fig.~\ref{fig:masterbestgraficomassa}).

The information criteria for the present case are:
\begin{equation}
AIC = 47.09,  \qquad BIC = 50.08.
\end{equation}

For the $\Lambda$CDM, using the results of \citet{fazzari2025effective}, where only $H_0$ is fitted:
\begin{align}
&     \chi^2_{\Lambda CDM} = 43.246,  \\
&AIC_{\Lambda CDM} = 45.35, \qquad BIC_{\Lambda CDM} = 46.24.
\end{align}

Hence,
\begin{equation}
\Delta AIC = -1.84, \qquad \Delta BIC = -3.83.
\end{equation}

These differences indicate that the two models are statistically comparable regarding the $AIC$. For the $BIC$, according to Tab.~\ref{tab:aic_bic_scale}, there is a moderate preference for the $\Lambda $CDM model. 
Although the $\Lambda$CDM is moderately favoured by the information criteria, the modified gravity scenario remains fully competitive.

Contrary to the binned Pantheon analysis, where the modified gravity model was strongly preferred, in the Master Sample the large number of free parameters increases the AIC/BIC. As a consequence, this increase penalizes the model despite the improvement in the chi-square.

\subsection{Joint conclusions from Pantheon and Master analyses}

The analysis performed on both the Pantheon Sample and the Master Sample shows that the model with an additional constraint represents a credible alternative to the standard $\Lambda$CDM cosmology. 
In both datasets, the model provides a good statistical description of the observations, as indicated by the $\Delta$AIC and $\Delta$BIC. Furthermore, the scalar degree of freedom introduced by the model is physically viable: its mass remains strictly positive and never diverges across the entire redshift range considered, a property consistently satisfied in both datasets.

The statistical performance of the present modified gravity model can be further evaluated by comparison with the power law (PL) model \citep{dainotti2021hubble, dainotti2022evolution}:

\begin{equation}
    \mathcal{H}(z)=\frac{H_0}{(1+z)^\alpha}.
    \label{PL}
\end{equation}
While the power law model interprets the effective running of $H_0$ as potentially astrophysical in origin, it can also be viewed as a phenomenological proxy for new physics \citep{schiavone2023f,montani2025two}.

To quantify the relative evidence in favor of the modified gravity scenario, the differences in Akaike and Bayesian information criteria are computed:
\begin{equation}
\Delta \mathrm{AIC} = \mathrm{AIC}_{PL} - \mathrm{AIC}_{\rm f(R)}, \qquad
\Delta \mathrm{BIC} = \mathrm{BIC}_{PL} - \mathrm{BIC}_{\rm f(R)}.
\end{equation}

The comparison for both the Pantheon and Master samples is summarized in Tab~\ref{tab:comparison_PL}.

\begin{table}[htbp]
\centering
\begin{tabular}{lcc}
\hline
Dataset & $\Delta \mathrm{AIC}$ & $\Delta \mathrm{BIC}$ \\
\hline
Pantheon & -0.02 & -0.03 \\
Master   & -2.18 & -3.17 \\
\hline
\end{tabular}
\caption{Comparison of the modified gravity model with the power law model. The values of the AIC and the BIC for the power law model are taken from \citet{dainotti2025new}.}
\label{tab:comparison_PL}
\end{table}

The comparison with the power law model shows that the model under consideration is essentially equivalent to the power law description when applied to the binned Pantheon sample. In contrast, for the Master sample, the power law model is weakly favored according to the $\Delta AIC$ and moderately favoured according to the $\Delta BIC$. This is a consequence of the smaller number of free parameters of the power law with respect to the modified gravity model under examination, which leads to more favorable values of both statistical criteria.

\subsection{Analytical reconstruction of \texorpdfstring{$f(R)$}{f(R)} in the low redshift limit}

In this model, the analytical reconstruction of an approximate low redshift form of $f(R)$ is achieved by combining the scalar-tensor representation with a Taylor expansion of the numerically computed background solution. The scalar field $\phi(z)$ and the potential $\bar{U}(z)\equiv\frac{U}{6H_0^2}$ are first determined numerically from the fit cosmological evolution. Both functions are then expanded around $z=0$ up to second-order, with numerical derivatives providing the expansion coefficients.

The scalar field expansion reads
\begin{equation}
    \phi(z) \simeq \phi_0 + \phi_1 z + \frac{1}{2} \phi_2 z^2 + \mathcal{O}(z^3),
\end{equation}
with coefficients
\begin{equation}
\phi_0 = 1.00007, \qquad 
\phi_1 = 0.08252, \qquad 
\phi_2 = -0.05214.
\end{equation}

Similarly, the potential is expanded around $\phi \simeq \phi_0$ as
\begin{equation}
    \bar{U}(\phi) \equiv \frac{U}{6H_0^2} \simeq 
    A_0 + A_1 (\phi - \phi_0) + A_2 (\phi - \phi_0)^2 
    + \mathcal{O}\!\left[(\phi - \phi_0)^3\right],
\end{equation}
with
\begin{equation}
A_0 = -0.11146, \qquad 
A_1 = 1.43204, \qquad 
A_2 = 9.34277.
\end{equation}

At this order, the Ricci scalar is related to the potential by
\begin{equation}
    R = \frac{d\bar{U}}{d\phi} 
    \simeq A_1 + 2A_2(\phi - \phi_0),
\end{equation}
which can be inverted directly:
\begin{equation}
    \phi(R) \simeq 
    \phi_0 + \frac{R - A_1}{2A_2}.
\end{equation}

The function $f(R)$ follows from the standard scalar-tensor relation,
\begin{equation}
    f(R) = R \, \phi(R) - \bar{U}(\phi(R)),
\end{equation}
and, by substituting the expansions and truncating at second-order, one obtains
\begin{equation}
    f(R) \simeq m^2 B_0 + B_1 R + B_2\frac{R^2}{m^2},
    \label{f(R)ricostruita}
\end{equation}
with coefficients
\begin{align}
    B_0 &= \frac{A_1^2}{4A_2} - A_0 = 0.166335, \\
    B_1 &= \phi_0 - \frac{A_1}{2A_2} = 0.92336, \\
    B_2 &= \frac{1}{4A_2} = 0.0267587.
\end{align}

\begin{figure}[ht]
    \centering
    \includegraphics[width=0.7\textwidth]{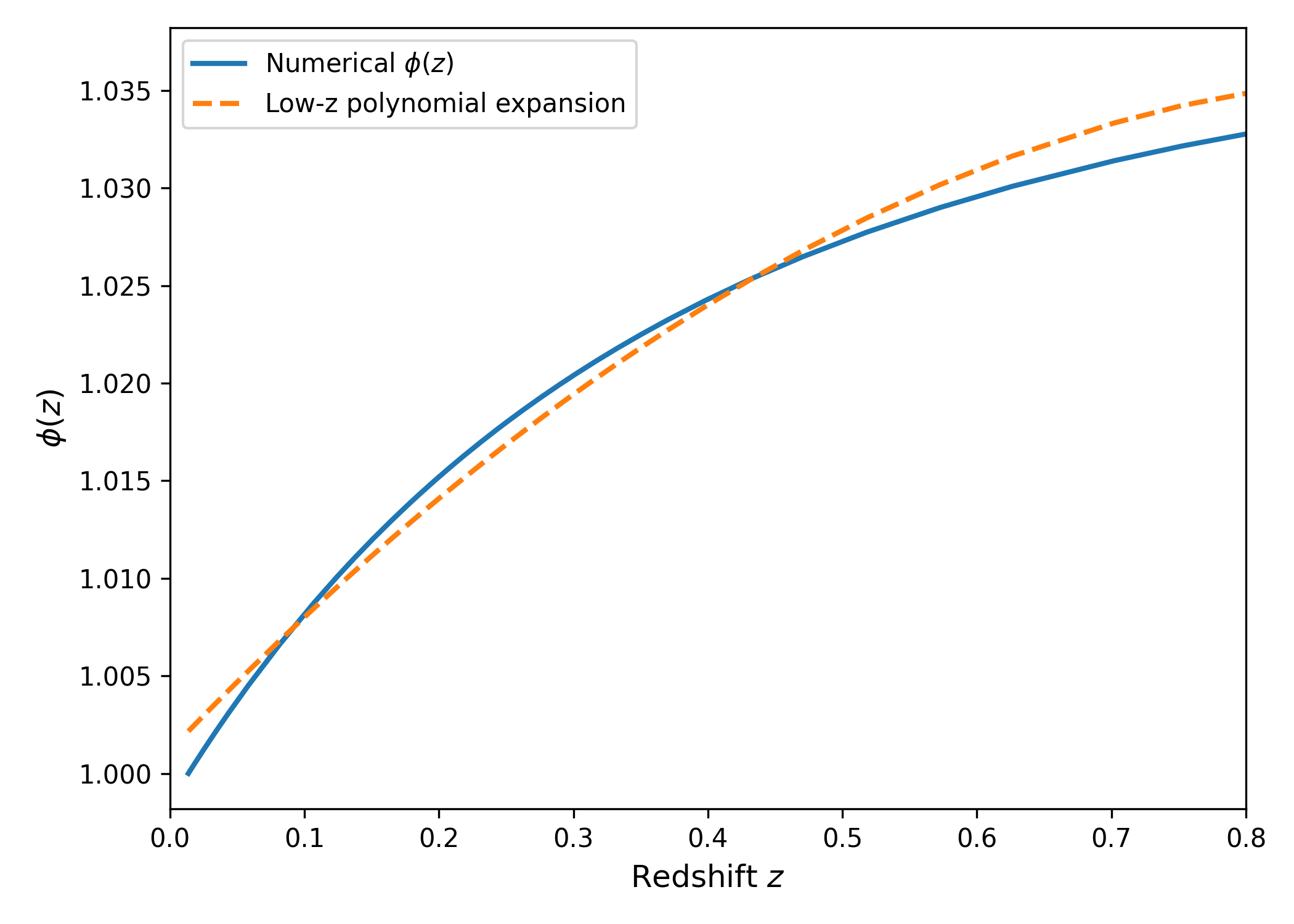}
    \caption{Comparison between the numerical solution of $\phi(z)$ and its low redshift quadratic expansion.}
    \label{fig:phi_expansion}
\end{figure}

\begin{figure}[ht]
    \centering
    \includegraphics[width=0.7\textwidth]{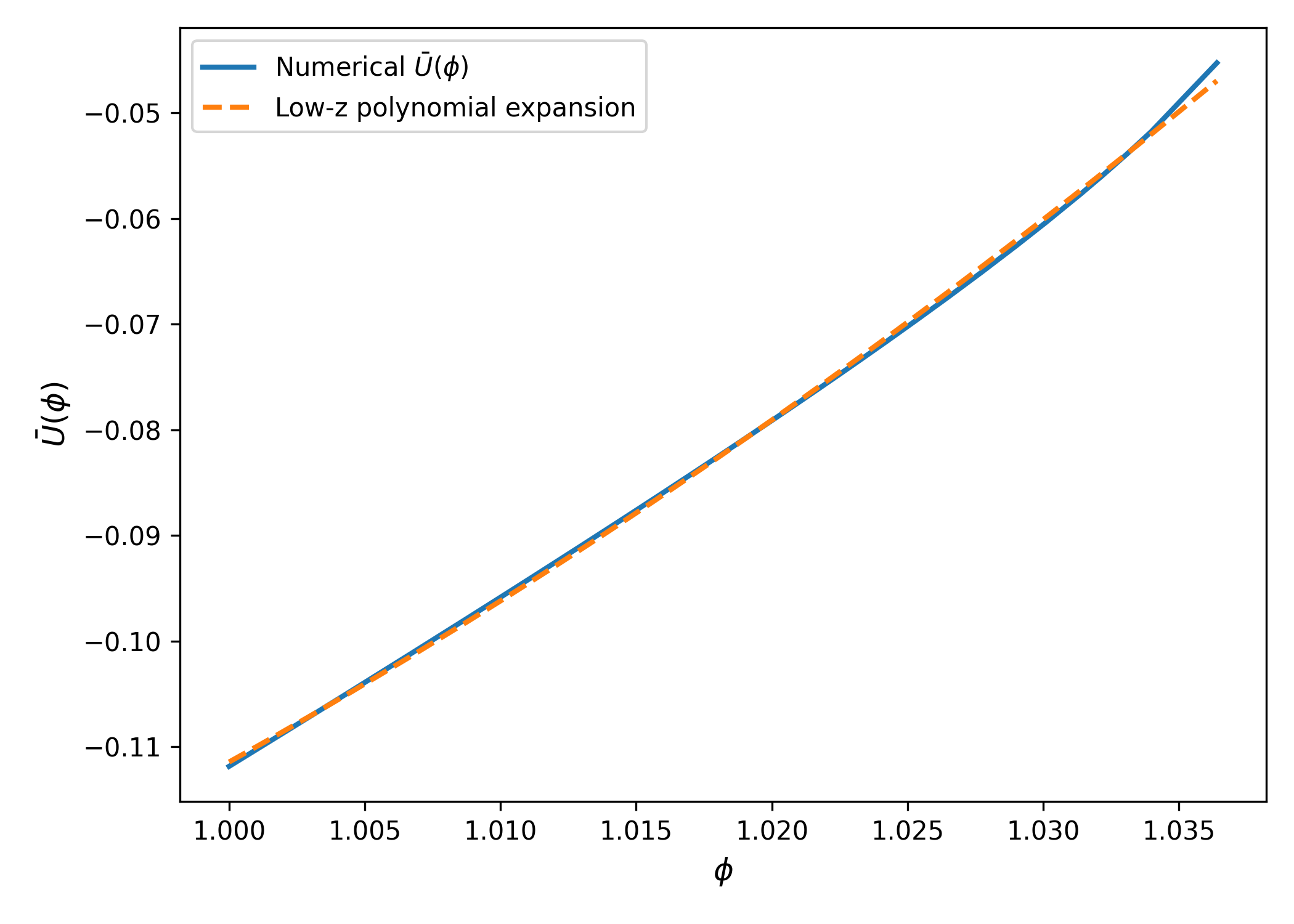}
    \caption{Comparison between the numerical potential $\bar{V}(\phi)$ and its quadratic expansion around $\phi_0$.}
    \label{fig:V_expansion}
\end{figure}

In Eq.~\eqref{f(R)ricostruita}, $m^2\equiv H_0^2$ is a characteristic curvature scale introduced to let the coefficients of the quadratic expansion of $f(R)$ be dimensionless and to set the scale of the theory.
To verify the accuracy of the reconstruction, the numerical solutions for $\phi(z)$ and $\bar{U}(\phi)$ are compared with their Taylor approximations for $z \lesssim 1.2$. Figures~\ref{fig:phi_expansion} and \ref{fig:V_expansion} show that the discrepancies never exceed $10^{-2}$, confirming that the quadratic expansions provide a sufficiently accurate representation in the low redshift regime.

The coefficient $B_0$ acts as an effective cosmological constant in the reconstructed theory. The coefficient $B_1$ is expected to be close to unity in order to recover the Einstein-Hilbert limit; the value $B_1 = 0.92336$ indicates a mild deviation from general relativity, as anticipated in a modified gravity model constructed to address cosmological tensions. 
The positivity of $B_2$ ensures the absence of tachyonic instabilities and therefore represents an important viability condition.

The coefficient $B_2$ is known to be strongly constrained by Solar System tests ($B_2 \lesssim 10^{-5}$, see \citet{bertotti2003test}). The reconstructed value does not satisfy such constraints. This discrepancy is expected, since the reconstructed $f(R)$ function is determined dynamically from cosmological data and is therefore valid only within the redshift range in which it has been constructed. It is not intended to be extrapolated to high-density environments such as the Solar System, in agreement with the chameleon mechanism, in which the effective mass strongly depends on the local matter density.  
In low-density cosmological environments the scalar field can remain light, affecting the dynamics at large scales, while in high-density regions it acquires a large mass, suppressing deviations from General Relativity.   
This mechanism has been extensively investigated in \citet{Brax2008,KhouryWeltman2004,Li2014}.

\section{Conclusions}
\label{conclusions}

In this work, two different formulations of metric $f(R)$ gravity were analysed. Their predictions for the effective running of the Hubble constant were compared with binned SNe Ia data from two distinct samples: the Pantheon sample \citep{Scolnic2018} and the Master sample \citep{dainotti2025new}.

The first $f(R)$ gravity model represents a general approach, in which a function of redshift appears directly in the denominator of the Hubble parameter. This function, which is specified \emph{a priori}, effectively determines the form of the gravitational Lagrangian once the reconstruction algorithm is applied. 
This function is chosen as a second-order Taylor expansion (Eq.\eqref{f(z)}), and the resulting model was fitted to both SNe Ia samples. Although the fitting procedure performs well, this formulation must be discarded, as it is invariably associated with problems in the scalar field mass.

The second formulation is based on a more specific assumption, in which the potential of the scalar field is treated as a dynamical quantity, allowing the form of $f(R)$ to be reconstructed \emph{a posteriori}. This approach provides a fit to the SNe Ia data of comparable quality to the $\Lambda$CDM, while ensuring full physical viability. In particular, the scalar field mass remains positive and finite.
The model under examination is statistically preferred over the standard cosmological one for the Pantheon Sample, while for the Master Sample the Bayesian evidence indicates compatibility with standard criteria \citep{verde2010statistical}.

The analysis further highlights the importance of the additional constraint introduced in the dynamical system, as discussed in previous works \citep{montani2024metric, montani2025decay}. In general formulations, such as the first model, the Cauchy problem for the non-minimally coupled scalar field is highly constrained: the field value must remain close to unity, and its initial time derivative must be compatible with zero. The additional \emph{ad hoc} condition restores freedom in the initial value of scalar field derivative.

The main result of this analysis is the identification and physical interpretation of the additional dynamical condition required for metric $f(R)$ gravity to remain viable when describing the effective running of the Hubble constant. 
It is found that this condition is not merely a phenomenological assumption, but a necessary ingredient to avoid an over-constrained scalar field dynamics and to ensure a finite and well-behaved scalar mass.

Although the specific dark energy sector associated with a given $f(R)$ function may determine whether the model provides a statistically better or worse description of the data compared to the $\Lambda$CDM scenario, the central achievement of this analysis lies in demonstrating the existence of a metric $f(R)$ model that can simultaneously explain the observational data and remain compatible with current theoretical and experimental constraints (see also \citet{NojiriOdintsovPhysRep}).

\paragraph{\textbf{Data Availability Statement}}
Data supporting the findings of this study are available upon request for the Pantheon and the Master sample.

\paragraph{\textbf{Code Availability Statement}}
The code used for the findings of this study is available upon request.

\paragraph{\textbf{CRediT statement}} 

\textbf{A. Valletta}: Conceptualization, Methodology, Formal analysis, Data analysis, Visualization, Writing -- original draft, Writing -- review \& editing.
\textbf{G. Montani}: Conceptualization, Methodology, Supervision, Writing -- original draft, Writing -- review \& editing.
\textbf{M. G. Dainotti}: Data curation, Formal analysis, Supervision, Writing -- review \& editing.
\textbf{E. Fazzari}: Software supervision, Writing -- review \& editing.

\section*{Acknowledgements}

M.G. Dainotti acknowledges the support of the DoS for her travel to La Sapienza in February 2025. M.G.D. acknowledges the support of the JSPS Grant-in-Aid for Scientific Research (KAKENHI) (A), Grant Number JP25H00675 for supporting her travel and accommodation to visit La Sapienza in September 2025.


\end{document}